\documentclass[prb,aps,showpacs]{revtex4}
\usepackage{graphicx}

\usepackage{bm}
\usepackage{color}

\newcommand{\beq}{\begin{equation}}
\newcommand{\beqa}{\begin{eqnarray}}
\newcommand{\eeq}{\end{equation}}
\newcommand{\eeqa}{\end{eqnarray}}

\begin{document}

\title{Spin relaxation in a generic two-dimensional
spin-orbit coupled system}

\author{Tudor D. Stanescu}
\affiliation{Condensed Matter Theory Center, Department of
Physics, University of Maryland, College Park, Maryland
20742-4111, USA}
\author{Victor Galitski}
\affiliation{Condensed Matter Theory Center and Joint Quantum
Institute, Department of Physics, University of Maryland, College
Park, Maryland 20742-4111, USA}
\date{\today}

\begin{abstract}
We study the relaxation of a spin density injected into a
two-dimensional electron system with generic spin-orbit
interactions. Our model includes the Rashba as well as linear and
cubic Dresselhaus terms. We explicitly derive a general
spin-charge coupled diffusion equation. Spin diffusion is
characterized by just two independent dimensionless parameters:
$\gamma_R$ and $\gamma_D$, which control the interplay between
different spin-orbit couplings. The real-time representation of
the diffuson matrix (Green's function of the diffusion equation)
is evaluated analytically. The diffuson describes space-time
dynamics of the injected spin distribution. We explicitly study
two regimes: The first regime corresponds to negligible
spin-charge coupling and is characterized by standard charge
diffusion decoupled from the spin dynamics. It is shown that there
exist several qualitatively different dynamic behaviors of the
spin density, which correspond to various domains in the
$(\gamma_R,\gamma_D)$ parameter space. We discuss in detail a few
interesting phenomena such as an enhancement of the spin
relaxation times, real space oscillatory dynamics, and anisotropic
transport. In the second regime, we include the effects of
spin-charge coupling. It is shown that the spin-charge coupling
leads to an enhancement of the effective charge diffusion
coefficient. We also find that in the case of strong spin-charge
coupling, the relaxation rates formally become complex and the
spin/charge dynamics is characterized by real time oscillations.
These effects are qualitatively similar to those observed in
spin-grating experiments [Weber et al., Nature  {\bf 437}, 1330
(2005)].
\end{abstract}

\maketitle

\section{Introduction}
Spin-orbit coupled systems host an amazing variety of interesting
effects,\cite{spintronQC} which are currently the subject of
intense experimental and theoretical investigation. The interest
in this field stems from the fact that spin-orbit coupling opens
the possibility of controlling the electron spin\cite{Loss} using
purely electrical
means\cite{Nitta,Grundler,Kato,Zhang_Science,Sinova_etal} with
potential applications in spintronics\cite{ZFDS,Wolf} and quantum
computing\cite{spintronQC}. The microscopic origin of the
spin-orbit interaction in a low-dimensional system is  the absence
of inversion symmetry in the semiconductor crystal and confining
potential. This leads to the appearance of the Rashba and
Dresselhaus (linear and cubic) spin-orbit terms in the effective
Hamiltonian.~\cite{Rashba,Rashba1,Dressel,DyakKach} The spin-orbit
interaction  results in a variety of macroscopic effects such as
antilocalization,\cite{Bergmann,Knap} electric-field induced spin
accumulation,\cite{Jedema,EHR} spin-galvanic
effects,\cite{Ganichev} etc.

One particularly important aspect of spin transport in disordered
systems is  Dyakonov-Perel spin relaxation.\cite{DyakPer} The
source of this effect is the precession of the electron spin
around a momentum-dependent magnetic field. The electron momentum
changes randomly due to electron scatterings off of impurities,
which in turn randomizes spin precession. One should note here
that a large spin relaxation rate is typically considered an
undesirable effect from the experimental point of view. On the one
hand,  the Dyakonov-Perel spin relaxation rate is proportional to
the strength of spin-orbit interactions. On the other hand, the
existence of relatively strong spin orbit-interaction is needed
for the appearance of non-trivial spin transport. It is therefore
important to understand the interplay between different phenomena
in spin transport and search for possibilities to reduce the
negative effects of spin relaxation.

We should mention here that the issue of spin relaxation
 has been extensively studied in experiments. In particular,
Weber et al. have used the transient spin grating method to study
spin dynamics in two-dimensional semiconductor
systems.\cite{Cameron,WeberAwschal,WeberAwschal2}. This
experimental technique is based on optical injection of spin
polarized electrons in a two-dimensional quantum well, which
generates a spin density wave with a wave-vector in the plane of
the two-dimensional system and the spin polarization perpendicular
to the plane. Particularly interesting results revealed by these
experiments\cite{WeberAwschal} were an enhancement of the charge
diffusion coefficient and an unusual time-dependence of the spin
grating amplitude, which exhibited oscillatory features in
addition to the conventional exponential decay.

Motivated by the aforementioned experimental results, we study
theoretically spin transport and spin relaxation in a
two-dimensional electron system with generic spin orbit
interaction. We focus our attention on the interplay between
different types of spin-orbit couplings and the effects of the
spin-charge coupling on the spin relaxation dynamics. We work
within the framework of the diffusion approximation, which assumes
that the spin relaxation length is large compared to the mean-free
path. We find that depending on the relative values of the Rashba,
linear and cubic Dresselhaus couplings one can have different
dynamic spin-relaxation regimes and phenomena such as oscillations
of the spin density in real space, anisotropic spin transport, and
enhancement of spin relaxation times. We also study in detail the
effect of spin-charge coupling on the diffusive dynamics. We find
that it always leads to a renormalization of the effective charge
diffusion coefficient, which gets enhanced compared to the
``bare'' diffusion constant. We also find that a strong
spin-charge coupling may formally lead to a complex
spin-relaxation rate and real-time oscillations in the spin and
charge diffusion behavior, which are qualitatively similar to the
oscillatory behavior observed in experiment. We should note here
that the real-time oscillations appear only beyond the formal
regime of validity of the diffusion approximation. However, they
give a tentative indication that the experimentally observed
oscillations may originate from the spin-charge coupling.

Our paper is structured as follows:

In Sec.~\ref{SecII}, we introduce our model Hamiltonian for the
non-interacting disordered two-dimensional electron system with
spin-orbit interaction and derive a general
 spin-charge coupled diffusion equation. We introduce natural
 dimensionless units of length and time and show that in these
 units, the spin part of the diffusion equation is characterized by
 two dimensionless parameters, $\gamma_R$ and $\gamma_D$. These
 parameters  are constrained to lie within a circle of radius two in the
 $(\gamma_R,\gamma_D)$ parameter space. We also introduce the
 matrix Green's function of the diffusion equation (diffuson),
 whose real time representation describes the spin relaxation
 dynamics of interest.

In Sec.~\ref{SecIII}, we consider the spin sector of the diffusion
equation (which is justified if the spin-charge coupling is
relatively small). We focus our attention on the diagonal
$z$-component of the diffuson, which describes the dynamics of
injected out-of-plane spin density. We show that there exist
different domains in the $(\gamma_R,\gamma_D)$-parameter space,
which correspond to qualitatively different dynamic regimes. We
explicitly derive the real space-time behavior of the diagonal
component of the diffuson matrix in various regimes. We also
discuss the spin dynamics as a function of momentum and derive the
spin relaxation spectrum as a function of spin-orbit coupling
parameters. In particular, we study in detail the ``symmetry
point'' in the parameter space ($\gamma_R=\gamma_D$), when the
relaxation time becomes infinite in certain directions. We study
the behavior of the spin relaxation spectrum  in the vicinity of
the symmetry point.

In Sec.~\ref{SecIV}, we discuss the effects of spin charge
coupling on the spin relaxation spectrum. We show that the spin
charge coupling may formally lead to a complex spin relaxation
rate, which physically translates into an oscillatory behavior of
the spin density. We discuss the validity of the diffusion
equation approximation and the implication of our results for
understanding spin transport experiments. Our conclusions and a
discussion of several open questions are presented in the last
Section~\ref{SecV}.

\section{The diffusion equation}\label{SecII}

In this section we introduce the model and derive a general form
of the spin-charge coupled diffusion equation. The main results of
this section are equations (\ref{Dijm1}) and (\ref{calDr})
representing the most general form of the diffusion equation in an
electron system with arbitrary spin-orbit interaction. In this and
subsequent Sections, we use the units $\hbar = k_{\rm B} = 1$.

We start with the following Hamiltonian, which describes the
conduction band electrons in a III-V type semiconductor quantum
well grown in the [001] direction (set as the z axis), \beq H =
\frac{{\bf p}^2}{2m} + {\bf h}({\bf p})\cdot \hat{\bf \sigma},
\label{ham} \eeq where $m$ is the effective electron mass,
$\hat{\bm \sigma} =(\hat{\sigma}_x,\hat{\sigma}_y, \hat{\sigma}_z
)$ is the Pauli matrix, and ${\bf h}({\bf p})=(h_x, h_y, h_z)$ are
functions of the two-dimensional momentum ${\bf p}$ describing the
spin-orbit interaction. We assume that the noninteracting
electrons are in the presence of a short-range impurity potential
and we investigate the role of the spin-orbit interaction in the
coupled spin and charge transport using the diffusion
approximation. In general, the second term of the Hamiltonian
(\ref{ham}), which describes the spin-orbit interaction, contains
both Rashba and Dresselhaus contributions. The Rashba
interaction\cite{Rashba,Rashba1}, which is due to the inversion
asymmetry of the quantum well confining potential, has the form
\beq {\bf h}^{R}({\bf p}) = \alpha v_F (p_y, -p_x), \eeq where
$v_F$ is the Fermi velocity, and $\alpha$ is a dimensionless
coupling constant. Throughout this work we will assume that the
spin-orbit coupling energy is much smaller than the Fermi energy,
i.e. $|\alpha|\ll 1$. Consequently, we will neglect the change of
the Fermi surface due to the presence of spin-orbit interaction.
 The Dresselhaus spin-orbit coupling,~\cite{Dressel} arising from the lack of inversion
symmetry of the semiconductor crystal, contains terms both linear
and cubic in ${\bf p}$, \beqa
{\bf h}^{D_1}({\bf p}) &=& \beta_1 v_F (p_x, -p_y), \\
{\bf h}^{D_3}({\bf p}) &=& -4\beta_3 \frac{v_F}{p_F^2}(p_xp_y^2,
-p_yp_x^2), \eeqa where $\beta_1$ and $\beta_3$ are dimensionless
coupling constants with $|\beta_j|\ll 1$, and $p_F=mv_F$ is the
Fermi momentum.

In the presence of disorder, the complete description of the
coupled spin and charge diffusive transport at long lengthscales
compared to the mean-free path is given by a set of linear
differential equations that, in its most general form, can be
written as \beq\left(\frac{\partial}{\partial t}-{\cal
D}{\bm\nabla}^2\right)\rho_i =
\left(-\Gamma^{ij}+P^{ijm}\nabla_m+{\bf C}^{ij}{\bm
\nabla}\right)\label{diffEq}\rho_j\eeq where $\partial/\partial t$
represents the time derivative, $\rho_0$ is the charge density,
while $\rho_1=\rho_x$,   $\rho_2=\rho_y$, and $\rho_3=\rho_z$ are
the densities of the corresponding spin components. In equation
(\ref{diffEq}) ${\cal D}$ represents the diffusion constant and
can be expressed in terms of the mean scattering time $\tau$ as
${\cal D}=\tau v_F^2/2$, $\Gamma^{ij}=(1/\tau_s)_{ij}$ describes
the Dyakonov-Perel spin relaxation\cite{DyakPer}, $P^{ijm}$  is
responsible for the precession of the inhomogeneous spin
polarization, and the last term, having non-zero elements ${\bf
C}^{i0}={\bf C}^{0i}$, describes the spin-orbit mixing of spin and
charge degrees of freedom. For convenience, throughout this
article we will express times in units of spin relaxation time,
$\tau_s$, and lengths in units spin relaxation length, $L_s$,
unless otherwise stated. The quantities  $\tau_s$ and $L_s$ depend
explicitly on the parameters of the 2D electron system, \beqa L_s
&=& \frac{1}{2 k_F [\alpha^2 +(\beta_1-\beta_3)^2
+\beta_3^2]^{1/2}}. \label{Ls}  \\ \tau_s &=&
 \frac{2\tau}{g^2[\alpha^2 +(\beta_1-\beta_3)^2
+\beta_3^2]}, \label{ts} \eeqa where $g=2v_Fk_F\tau$ is a
coefficient proportional to the dimensionless conductance. Using
the units (\ref{Ls},\ref{ts}) , we can write the coefficients from
Eq.~(\ref{diffEq}) as dimensionless coupling constants. For
example, the diagonal contributions to the spin relaxation terms
are $\Gamma_{xx}=\Gamma_{yy}=1$ and $\Gamma_{zz}=2$. The final
matter of convenience concerns the parametrization of the
spin-orbit interaction. This interaction can be described in terms
of the original coupling constants $(\alpha, \beta_1, \beta_3)$
or, alternatively, in terms of a new set of parameters, $(\Gamma,
\gamma_R, \gamma_D)$, defined by \beqa \Gamma^2 &=&
\alpha^2+\beta_1^2+\beta_3^2 \label{gammas} \\
\gamma_R&=&\frac{2\alpha}{\sqrt{\alpha^2 +(\beta_1-\beta_3)^2
+\beta_3^2}}, ~~~~~~~~~~~~~
\gamma_D=\frac{2(\beta_1-\beta_3)}{\sqrt{\alpha^2
+(\beta_1-\beta_3)^2 +\beta_3^2}}. \nonumber \eeqa The parameter
$\Gamma$ represents a measure of the overall strength of the
spin-orbit interaction, while $\gamma_R$ and $\gamma_D$
characterize the interplay between the Rashba and Dresselhaus
contributions. By convention the sign of $\Gamma$ is the same as
the sign of $\beta_3$. We derive the diffusion equations using the
standard density matrix formalism\cite{Mahan}, used previously in
 Refs. [\onlinecite{BurkovNunMcD,Malshuk,Malsh2,Shyt}].
Eq.~(\ref{diffEq}) can by re-written in the form \beq \left[
\delta_{ij} - {\Pi}_{ij} \right]~\rho_j=0, \label{PiDef} \eeq
where, for a homogeneous system, $\delta_{ij} -
{\Pi}_{ij}(\omega,{\bf k}) =
(-i\omega+k^2)\delta_{ij}+\Gamma^{ij}+P^{ijm}k_m+{\bf C}^{ij}{\bf
k}$ and summations over repeated indices are  assumed. The matrix
$\hat{\Pi}$ represents the kernel of the diffusion equation and
can be expressed in terms of retarded and advanced Green functions
as \beq \hat{\Pi}_{ij}(\omega, {\bf k}) =\frac{1}{4\pi\nu_F\tau}
\int \frac{d^2{\bf q}}{\left( 2 \pi \right)^2}
\mbox{Tr}\left\{\hat{\sigma}_i~\hat{G}^R(\omega/2,{\bf k}+{\bf
q}/2)~\hat{\sigma}_j~\hat{G}^A(\omega/2,{\bf k}-{\bf
q}/2)\right\}, \label{PiGG}\eeq where $\hat{G}^{R(A)}$ is the
$2\times 2$ retarded (advanced) Green function matrix averaged
over disorder, $\hat{\sigma}_i$ are the Pauli matrices, $\nu_F$ is
the density of states at the Fermi energy and $\mbox{Tr}\{\dots\}$
involves the trace over spin indices. The diffusion approximation
represents the low frequency, long wavelength limit of
Eq.~(\ref{PiGG}). This approximation is valid as long as the
scattering time is much shorter than the spin precession time,
i.e. in the weak spin-orbit coupling limit $\Gamma v_F k_F \tau
\ll 1$. Taking the appropriate small frequency and small momentum
limit, we obtain \beq \hat{1}-\hat{\Pi}(\omega, {\bf k}) = \left(
\begin{array}{llll}
~~~~~~~~~~s-1 ~~~~~~~~-ig(\mu_0 k_x + \mu_1 k_y)~~~~ ig(\mu_1 k_x + \mu_0 k_y)~~~~~~~~~~~~~0  \\
-ig(\mu_0 k_x + \mu_1 k_y) ~~~~~~~~~~~~ s  ~~~~~~~~~~~~~~~~~~~ \frac{1}{2}\gamma_R\gamma_D~~~~~~~~~ -i(\gamma_R k_x+\gamma_D k_y)\\
~~ig(\mu_1 k_x + \mu_0 k_y) ~~~~~~~~~ \frac{1}{2}\gamma_R\gamma_D  ~~~~~~~~~~~~~~~~~~~~ s~~~~~~~~~~~-i(\gamma_D k_x+\gamma_R k_y)\\
~~~~~~~~~~~~~0~~~~~~~~~~~~~~i(\gamma_R k_x+\gamma_D
k_y)~~~~i(\gamma_D k_x+\gamma_R k_y)~~~~~~~~~~~~ s+1
\end{array}    \right),  \label{Dijm1}  \eeq
where $\hat{1}$ is the $4\times 4$ unit matrix, $s=-i\omega+k^2+1$
and the spin-charge coupling is characterized by the parameters
\beqa \mu_0 &=&
\frac{(3\beta_3-\beta_1)(\alpha^2-\beta_1^2+\beta_3^2)-\beta_1\beta_3^3}{\sqrt{\alpha^2
+
(\beta_1-\beta_3)^2 +\beta_3^2}},  \nonumber \\
\mu_1 &=&
\frac{\alpha(\alpha^2-\beta_1^2+6\beta_3^2)}{\sqrt{\alpha^2
+(\beta_1-\beta_3)^2 +\beta_3^2}}. \label{mus}
 \eeqa
In addition, it will be  convenient to consider the diffusion
problem within  a coordinate system that is rotated in the x-y
plane counterclockwise with $\pi/2$ relative to the initial
coordinate system\cite{Bernev}. The rotated coordinates are $r_+
=(y+x)/\sqrt{2}$ and $r_- = (y-x)/\sqrt{2}$, while the new
components of the spin density will be $\rho_+ =
(\rho_y+\rho_x)/\sqrt{2}$ and $\rho_- =(\rho_y-\rho_x)/\sqrt{2}$,
in addition to $\rho_z$ and the charge density $\rho_0$ that
remain unchanged. Relative to the rotated coordinate system,
diffusion is described by the operator \beq \hat{1} -
\hat{\Pi}^{(r)}(\omega, {\bf k}) = \left(\begin{array}{cccc}
s-1 & i\xi(\mu_0- \mu_1)k_- & i\xi(\mu_0 + \mu_1)k_+ & 0 \\
i\xi(\mu_0- \mu_1)k_- & s +\frac{1}{2}\gamma_R\gamma_D & 0 & -i(\gamma_R+\gamma_D)k_+\\
i\xi(\mu_0 + \mu_1)k_+ & 0 & s-\frac{1}{2}\gamma_R\gamma_D & -i(\gamma_R-\gamma_D)k_-\\
0 & i(\gamma_R+\gamma_D)k_+ & i(\gamma_R-\gamma_D)k_- & s+1
\end{array}
\right).                  \label{calDr} \eeq where
 the four columns correspond to $\rho_0$, $\rho_+$,  $\rho_-$, and $\rho_z$,
respectively. Equations (\ref{Dijm1}) and (\ref{calDr}) are the
main results of this section.

The physical problem that we will be solving on the basis of
Eqs.~(\ref{Dijm1},~\ref{calDr}) is to describe the transport of an
injected spin density in the presence of a general type of
spin-orbit coupling. Assuming that a spin (or charge) density with
components $\rho_i(0, {\bf r})$ was injected into the system at
$t=0$, the density profile at times $t>0$ will be \beq \rho_i(t,
{\bf r}) = \int dr^{\prime}~D_{ij}(t,{\bf r}, {\bf
r}^{\prime})\rho_j(0,{\bf r}^{\prime}), \eeq where $\hat{D}$ is
the diffuson, i.e., the Green's function of the diffusion
equation. The physical meaning of the diffuson can be understood
if we assume that initially we inject a $\delta$-like, density
profile $\rho_j(0, {\bf r})= \delta({\bf r})$. The spin and charge
densities at any later time are given by the matrix elements of
the diffuson, $\rho_i(t, {\bf r}) = D_{ij}(t, {\bf r})$. For a
homogeneous system the diffuson is the inverse of the diffusion
kernel,
\begin{equation}
\label{diffuson} \hat{D}=\left[ \hat{1} - \hat{\Pi}\right]^{-1}.
\end{equation}
Consequently, solving the transport problem for an injected spin
density implies inverting the matrix in Eq.~(\ref{Dijm1}) or
(\ref{calDr}) and performing the appropriate Fourier transforms.
We notice that the spin and charge dynamics is controlled by the
poles of $\hat{D}(\omega, {\bf k})$. These poles determine  four
relaxation modes $i\omega_{\alpha}({\bf k})=1/\tau_{\alpha}({\bf
k})$, which follow from the equation:
\begin{equation}
\label{eigen} \Delta(\omega,{\bf k}) \equiv \mbox{det}[\hat{1} -
\hat{\Pi}(s, {\bf k})] = 0
\end{equation}
A time dependent matrix element of the diffuson can be written in
general as \beq \mbox{D}_{ij}(t, {\bf k}) = \sum_{l=0}^3 A_l({\bf
k}) e^{-i\omega_l({\bf k})~t}, \label{genSol}\eeq where the
amplitudes $A_l({\bf k})$ are functions of momentum. We should
emphasize here that the long time relaxation of the injected spin
density is uniquely determined by the lowest minimum of the
relaxation rate modes. In a standard diffusion problem, the charge
relaxation rate is $1/\tau_0({\bf k})={\cal D} k^2$ (where ${\cal
D}$ is the diffusion constant, which is equal to one in our
special units) and the minimum obviously corresponds to ${\bf k}=
{\bf 0}$. By contrast, in the presence of spin-orbit interaction,
the k-dependence of the relaxation rate is more complicated, with
possible minima at finite momenta. The detailed dependence of the
relaxation rate on the spin-orbit interaction is analyzed in the
following sections.

\section{The solution of the time-dependent diffusion equation: The spin
sector}\label{SecIII}

In this section we discuss the spin relaxation spectrum and derive
the diffuson in momentum space~\ref{kspace} and real
space~\ref{rspace}. The main results of section \ref{kspace} are
Eqs. (\ref{iwj}, \ref{iw0kapp}), describing the  spin relaxation
spectrum and Fig.~\ref{FIGsr1}, which summarizes the properties of
the spectrum. The behavior in the vicinity of the symmetry points
characterized by an infinite spin relaxation time is illustrated
in Fig.~\ref{FIGsr2}. Eqs.~(\ref{Disotrop}) and
(\ref{redasymm}-\ref{Dtr2k}) describe spin diffusion in real
space.

\subsection{Momentum space picture}
\label{kspace}

\begin{figure}
\begin{center}
\includegraphics[width=0.4\textwidth]{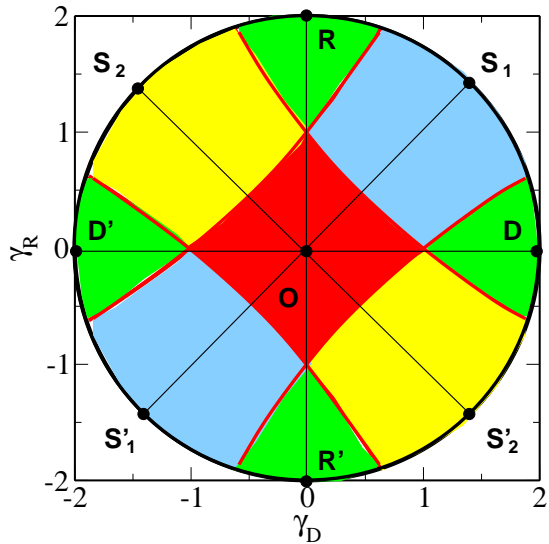}
\caption{Diagram summarizing the the types of minima
characterizing the lowest energy relaxation rate modes $i\omega_1$
and $i\omega_2$ (see Eq.~(\ref{iwj})). The spin sector is uniquely
described by the dimensionless spin-orbit coupling constants
$\gamma_R$ and $\gamma_D$. These parameters are proportional to
the Rashba interaction and to the difference between the linear
and cubic Dresselhaus couplings, respectively, and they satisfy
the physical constraint $\gamma_R^2+\gamma_D^2\leq4$ (see
Eq.~(\ref{gammas})). In the red zone, both modes $i\omega_1$ and
$i\omega_2$ have the minimum at ${\bf k}=0$. In the blue (yellow)
regions, the minimum of the $i\omega_1$ ($i\omega_2$) mode is at
zero momentum, while the minimum of the $i\omega_2$ ($i\omega_1$)
mode is at a finite wave-vector $k_+^0$ ($k_-^0$) given by
Eq.~(\ref{MINk}). In the green zones, both modes have a finite
k-vector minimum at $k_{-}^0$ (for $i\omega_1$) and $k_{+}^0$ (for
$i\omega_2$). The points R and R' correspond to the the pure
Rashba case $(\gamma_R, ~\gamma_D)=(\pm2, ~0)$, while the segment
DD' corresponds to the pure Dresselhaus spin-orbit interaction,
$\gamma_R=0$, $\gamma_D\in[-2,2]$. At the end points
$\gamma_D=\pm2$ the cubic Dresselhaus term vanishes. The  spin
dynamics is isotropic for $\gamma_R\gamma_D=0$, i.e. along the RR'
and D'D segments. $S_j$ and $S_j^{\prime}$ are special symmetry
points characterized by a zero minimum relaxation rate
$i\omega_j(k^0)$, i.e. an infinite spin relaxation time.}
\label{FIGsr1}
\end{center}
\end{figure}

The diffusion approximation leading to equation (\ref{Dijm1}) is
rigourously valid only in the weak spin-orbit coupling regime
characterized by $\Gamma v_F k_F \tau \ll 1$. In this limit, the
spin-charge coupling is small, $g|\mu_{1,2}|\ll1$,  and can be
neglected in the leading approximation. The goal of this section
is to determine the role played by the interference between
different spin-orbit interaction terms in spin diffusion. In
particular, motivated by experiments that probe spin dynamics
optically, we concentrate on the properties of the out-of-plane
component of the spin density. To address this problem, we have to
determine the element ${D}_{zz}=[\hat{1} - \hat{\Pi}^{-1}]_{zz}$
of the Green's function for the diffusion equation  (the
diffuson). The general expression for the diagonal z-component of
the diffuson in the absence of spin-charge coupling is
$$
{D}_{zz}(s, {\bf k}) = [s^2-(\gamma_R\gamma_D/2)^2]/\Delta(s, {\bf
k}),
$$
where $\Delta(s, {\bf k})$ is a third order polynomial in $s$,
\begin{eqnarray}
\Delta(s, {\bf k}) &=&
s^2(s+1)-s\left[\left(\gamma_R^2+\gamma_D^2\right)k^2 +
4\gamma_R\gamma_D k_x k_y +
\left(\frac{\gamma_R\gamma_D}{2}\right)^2\right] \nonumber \\
&+&\gamma_R^2\gamma_D^2k^2 +
\gamma_R\gamma_D\left(\gamma_R^2+\gamma_D^2\right)k_x k_y
-\left(\frac{\gamma_R\gamma_D}{2}\right)^2
\end{eqnarray}
The spin dynamics is determined by the three spin relaxation rate
modes $i\omega_j({\bf k}) = -s_j({\bf k})+k^2+1$, obtained by
solving the equation $\Delta(s, {\bf k})=0$. For later reference,
let us identify the modes that control the spin dynamics, together
with  the charge mode,  according to their zero momentum values,
\beqa i\omega_0(0)
&=& 0, \nonumber \\
i\omega_1(0) &=& 1-\frac{\gamma_R\gamma_D}{2}, \nonumber \\
i\omega_2(0) &=& 1+\frac{\gamma_R\gamma_D}{2}, \label{modes} \\
i\omega_3(0) &=& 2, \nonumber \eeqa  where the mode $i\omega_0$ is
responsible for charge diffusion and has the standard form
$i\omega_0({\bf k})=k^2$ if we neglect spin-charge coupling.  Let
us first focus on the experimentally relevant special case of
momentum parallel to the [110] or [1$\bar{1}$0] directions
corresponding to $k_{\pm}$. Inverting the matrix from
Eq.~(\ref{calDr}) we obtain, \beq {D}_{zz}(s, k_{\pm}) =
\frac{s\pm\frac{\gamma_R\gamma_D}{2}}{s^2+s\left(1\pm\frac{\gamma_R\gamma_D}{2}\right)-k_{\pm}^2(\gamma_R\pm\gamma_D)^2\pm
\frac{\gamma_R\gamma_D}{2}}. \label{Dzzsk} \eeq Notice that, for
k-vectors oriented along these special directions, one of the
modes $i\omega_1$ or $i\omega_2$ does not contribute to the
dynamics and, consequently, the problem simplifies significantly.
The time dependence of $D_{zz}$ can be obtained by Fourier
transforming (\ref{Dzzsk}) with respect to the frequency. This
time dependence is uniquely determined by the roots of the secular
equation $\Delta(s, k_{\pm})=0$, i.e. by \beqa i\omega_{1}(k_{-})
&=& \frac{3}{2} + k_{-}^2 -\frac{\gamma_R\gamma_D}{4}
-\frac{1}{2}\sqrt{\left(1+\frac{\gamma_R\gamma_D}{2}\right)^2+4k_{-}^2(\gamma_R-\gamma_D)^2},
\nonumber \\
i\omega_{2}(k_{+}) &=& \frac{3}{2} + k_{+}^2
+\frac{\gamma_R\gamma_D}{4}
-\frac{1}{2}\sqrt{\left(1-\frac{\gamma_R\gamma_D}{2}\right)^2+4k_{+}^2(\gamma_R+\gamma_D)^2},
\label{iwj} \\
i\omega_{3}(k_{\pm}) &=& \frac{3}{2} + k_{\pm}^2
\pm\frac{\gamma_R\gamma_D}{4}
+\frac{1}{2}\sqrt{\left(1\mp\frac{\gamma_R\gamma_D}{2}\right)^2+4k_{\pm}^2(\gamma_R\pm\gamma_D)^2}.\nonumber
 \eeqa
 As mentioned above,
the modes $i\omega_1(k_+)=1+k_+^2-\gamma_R\gamma_D/2$ and
$i\omega_2(k_-)=1+k_-^2+\gamma_R\gamma_D/2$ do not contribute to
the dynamics  of the out-of-plane component of the
 spin.  The explicit time dependence of $D_{zz}$ for k-vectors along the '+'
 direction is \beq
{D}_{zz}(t, k_{+}) =
 \frac{1}{2}\left[e^{-i\omega_2(k_{+})t}+e^{-i\omega_3(k_{+})t}\right]
 -\frac{\left(1-\frac{\gamma_R\gamma_D}{2}\right)}
 {2\sqrt{\left(1-\frac{\gamma_R\gamma_D}{2}\right)^2+4k_{+}^2(\gamma_R+\gamma_D)^2}}
\left[e^{-i\omega_2(k_{+})t}-e^{-i\omega_3(k_{+})t}\right].
\label{DzztkGen}
 \eeq
 A similar expression can be obtained for ${D}_{zz}(t, k_-)$ by
 replacing $k_+$ with $k_-$, $i\omega_2$ with $i\omega_1$ and changing the sign
 of $\gamma_D$. The spin dynamics at large times, $t\gg\tau_s$, is
controlled by the minima of the relaxation rate modes
$i\omega_j(k_{\pm})$. The mode $i\omega_3$ has always the minimum
at $k=0$, $i\omega_2(0) = 2$. On the other hand, the position and
the values of the minima for the other two modes depend on the
spin-orbit coupling parameters. The mode $i\omega_1$ has a minimum
$i\omega_1(0) = 1-\gamma_R\gamma_D/2$ at zero wave-vector if
$\gamma_R^2+\gamma_D^2-5\gamma_R\gamma_D/2<1$. Otherwise, the
dispersion curve has a local maximum at $k=0$ and the minimum
occurs at a finite wave-vector $k_-^0$. Similarly, the mode
$i\omega_2$ has either a zero wave-vector minimum, if
$\gamma_R^2+\gamma_D^2+5\gamma_R\gamma_D/2<1$, or otherwise a
finite k-vector minimum at $k_+^0$. The finite wave-vectors that
describe the positions of the minima are \beq |k_{\pm}^0|
=\frac{\sqrt{(\gamma_R\pm\gamma_D)^4-\left(1\mp\frac{\gamma_R\gamma_D}{2}\right)^2}}{2|\gamma_R\pm\gamma_D|},
\label{MINk} \eeq while the frequencies corresponding to the
minima in the dispersion curves are \beqa i\omega_1(k_{-}^0) &=&
\frac{3}{2}-\frac{\gamma_R\gamma_D}{4}-\frac{\left(1+\frac{\gamma_R\gamma_D}{2}\right)^2}{4(\gamma_R-\gamma_D)^2}
-\frac{(\gamma_R-\gamma_D)^2}{4} \nonumber\\ i\omega_2(k_{+}^0)&=&
\frac{3}{2}+\frac{\gamma_R\gamma_D}{4}-\frac{\left(1-\frac{\gamma_R\gamma_D}{2}\right)^2}{4(\gamma_R+\gamma_D)^2}
-\frac{(\gamma_R+\gamma_D)^2}{4}. \label{iwMINk}\eeqa A summary of
the conditions necessary for the existence of these minima in
terms of the spin-orbit coupling parameters $\gamma_R$ and
$\gamma_D$ is presented in Fig. \ref{FIGsr1}. In the red domain
the condition $\gamma_R^2+\gamma_D^2\pm5\gamma_R\gamma_D/2<1$ is
satisfied and all the dispersion curves have minima at zero
momentum. The blue regions are characterized by a finite
wave-vector minimum of the $i\omega_2$ mode along the $k_+$
direction and a zero momentum minimum of the $i\omega_1$ mode,
while the yellow sectors correspond to the [$i\omega_1(k_-^0)$,
$i\omega_1(0)$] pair of minima. Finally, in the green regions both
modes, $i\omega_1$ and $i\omega_2$ have finite k-vector minima
corresponding to $k_-^0$ and $k_+^0$, respectively.
Notice that spin diffusion is isotropic only if
$\gamma_R\gamma_D=0$, i.e. for pure Rashba spin-orbit interaction
($\beta_1=\beta_3=0$ or $(\gamma_R, \gamma_D) = (\pm2, 0)$, the
points R and R' in Fig. \ref{FIGsr1}), pure Dresselhaus spin-orbit
coupling ($\alpha=0$ or $\gamma_R=0$, the segment D'D in Fig.
\ref{FIGsr1}), or a combination of Rasba and Dresselhaus
contributions with the linear and cubic Dresselhaus terms being
equal ($\beta_1=\beta_3$ or $\gamma_D=0$ and $\alpha\neq 0$, the
segment RR' without the end points). In all the other cases the
spin diffuses anisotropically as a result of the interference
between the spin-orbit coupling terms.

To be more specific, let us  consider now the case $\gamma_R>0$,
$\gamma_D>0$, i.e. the quarter defined by the points R, O, and D
in Fig. \ref{FIGsr1}. We assume that a uniform out-of-plane spin
density was injected into the system and ask about its time
evolution. The answer can be obtained using  Eq.~(\ref{DzztkGen})
for $k_+=0$. For a uniform system, the  out-of-plane spin
diffusion equation that takes the trivial form $D_{zz}(t, 0) =
\exp(-2 t)$, which corresponds to $i\omega_3(0) = 2$ representing
the uniform spin relaxation rate of the $S_z$-component. If we
consider now the limit of small non-zero momenta an expand all the
quantities up to second order terms in k, we obtain \beq D_{zz}(t,
{\bf k}) \approx
\frac{(\gamma_R-\gamma_D)^2}{\left(1+\frac{\gamma_R\gamma_D}{2}\right)^2}~k_-^2e^{-i\omega_1({\bf
k})t} +
\frac{(\gamma_R+\gamma_D)^2}{\left(1-\frac{\gamma_R\gamma_D}{2}\right)^2}~k_+^2e^{-i\omega_2({\bf
k})t} + e^{-i\omega_3({\bf k})t}, \label{Dzz0kapp} \eeq where the
dispersion relations at small k-vectors take the form \beqa
i\omega_1({\bf k}) &\approx& i\omega_1(0) +
k^2-\frac{(\gamma_R-\gamma_D)^2}{1+\frac{\gamma_R\gamma_D}{2}}~k_-^2,
\nonumber \\  i\omega_2({\bf k}) &\approx& i\omega_2(0) +
k^2-\frac{(\gamma_R+\gamma_D)^2}{1-\frac{\gamma_R\gamma_D}{2}}~k_+^2,
\label{iw0kapp}\\  i\omega_3({\bf k}) &\approx& i\omega_3(0) +
k^2+\frac{(\gamma_R-\gamma_D)^2}{1+\frac{\gamma_R\gamma_D}{2}}~k_-^2
+
\frac{(\gamma_R+\gamma_D)^2}{1-\frac{\gamma_R\gamma_D}{2}}~k_+^2,
\nonumber \eeqa where $k^2=k_x^2+k_y^2=k_+^2+k_-^2$. The initial
decay of the spin density is dominated by the third term in
Eq.~(\ref{Dzz0kapp}) that has the form $\exp[-(2 +{\cal
D}_+k_+^2+{\cal D}_-K_-^2)t]$. Here, ${\cal D}_{\pm}$ are
effective diffusion constants along the $k_{\pm}$ directions. This
result shows explicitly that the diffusion of the out-of-plane
component of the spin is, in general, anisotropic, and that the
anisotropy increases rapidly as we approach the symmetry point
$S_1$. Another consequence of these relations is that the
asymptotic behavior at $t\gg\tau_s$ is controlled by the
$i\omega_1$ mode, unlike the uniform case described solely by
$i\omega_3$. In the isotropic case, the relaxation times
corresponding to these modes differ by a factor of two,
$\tau_1/\tau_3=i\omega_3/i\omega_1=2$. As we approach the symmetry
point, this ratio diverges,
$\tau_1/\tau_3=4/(2-\gamma_R\gamma_D)$. At the same time, the
pre-factor of the first term in Eq.~(\ref{Dzz0kapp}) vanishes as
we approach the $OS_1$ line (see Fig. \ref{FIGsr1}) characterized
by equal Rashba and Dresselhaus couplings, or if the wave-vector
is oriented along the $k_+$ direction. We conclude that in general
the asymptotic spin relaxation at small (but not-vanishing)
k-vectors differs strongly from the spin relaxation at $k=0$. In
addition, the small wave-vector behavior is, in general,
anisotropic as a result of the interference between different
spin-orbit coupling terms.

Next, we focus on the finite momentum minima of the dispersion
relations (\ref{MINk}) and (\ref{iwMINk}). The existence of these
minima for the $i\omega_1$ and $i\omega_2$ modes implies an
increase in the spin lifetime relative to $k=0$ for the
corresponding mode by a factor
$i\omega_{1(2)}(0)/i\omega_{1(2)}(k_{\mp}^0)$. For example, in the
pure Rashba case we obtain $k_{\pm}^0=\sqrt{15}/4$ in units of
$1/L_s$, or $k_{\pm}^0=\alpha k_F\sqrt{15}/2$ in standard units,
and an increase in the lifetime by a factor of 16/7, in agreement
with previous calculation\cite{BurkovNunMcD,Frolts} based on the
Rashba model.

\begin{figure}
\begin{center}
\includegraphics[width=0.6\textwidth]{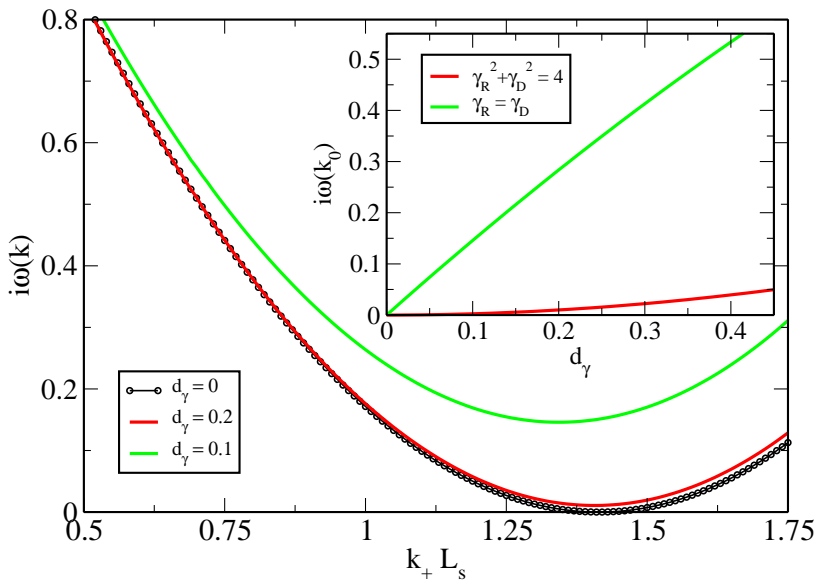}
\caption{Momentum dependence of the relaxation rate mode
$i\omega_2(k_+)$ for spin-orbit coupling parameters in the
vicinity of the symmetry point $\gamma_R=\gamma_D=\sqrt{2}$ (point
$S_1$ in Fig. \ref{FIGsr1}). In this regime, the $i\omega_2$ mode
has a finite wave-vector minimum along the $k_+$ direction. At the
symmetry point, the dispersion curve becomes gapless (black curve
and circles), meaning that the spin relaxation time at
$k_+^0=\sqrt{2}/L_s$ is infinite. As we depart from the symmetry
point, a finite gap opens. The spin-orbit coupling parameters for
the red curve correspond to a point on the $RS_1D$ arc in Fig.
\ref{FIGsr1}, while the green curve is for a point on the $S_1O$
segment, both points being in the vicinity of $S_1$.
 The gap size as a function of the distance $d_{\gamma}$ from the symmetry
 point [see Eq.~(\ref{d}) for the definition of $d_{\gamma}$] is shown in the inset.
 The red curve correspond to a spin orbit interaction with vanishing cubic
Dresselhaus contribution $\beta_3=0$ (the $RS_1D$ arc), while the
green curve was obtained for $\gamma_R=\gamma_D<\sqrt{2}$ (the
$S_1O$ segment). To observe a substantial enhancement of the spin
relaxation time at finite momentum, i.e. to have a small gap in
the relaxation rate curve, the cubic Dresselhaus spin-orbit
interaction $\beta_3$ should be minimal.} \label{FIGsr2}
\end{center}
\end{figure}

Another important special case is represented by the symmetry
points characterized by equal Rashba and Dresselhaus couplings,
more precisely defined by $|\gamma_R|=|\gamma_D|=\sqrt{2}$ (the
$S_j$ points in Fig. \ref{FIGsr1}). At a symmetry point the
minimum in the dispersion curve vanishes,
$i\omega_{2(1)}(\sqrt{2})=0$, reflecting the absence of spin
relaxation\cite{AverkGol, Bernev} at the corresponding
wave-vector, $k_+^0=\sqrt{2}$ (or $k_-^0=\sqrt{2}$ if
$\gamma_R\gamma_D<0$). The existence of a finite k-vector minimum
in the dispersion curves is illustrated in Fig. \ref{FIGsr2}. The
black curve with circles corresponds to the symmetry point
$\gamma_R=\gamma_D=\sqrt{2}$ and has a gapless minimum at
$k_+^0=\sqrt{2}/L_s$. As we move away from the symmetry point, a
gap develops (red and green lines). How is the gap behaving as we
vary the coupling parameters in the vicinity of the symmetry
point? To answer this question, we define $d_{\gamma}$ as the
distance in the parameter space $(\gamma_R, \gamma_D)$
 from the symmetry point
\begin{equation}
\label{d}
 d_{\gamma} =
[(\gamma_R-\sqrt{2})^2+(\gamma_D-\sqrt{2})^2]^{1/2}.
\end{equation}
 If we move away from the symmetry point by changing
the ratio between the Rashba and linear Dresselhaus interactions
in the absence of a cubic Dressselhaus term, i.e.  by varying the
ratio $\alpha/\beta_1$ and keeping $\beta_3=0$ (see the inset of
Fig. \ref{FIGsr2}, red curve), the gap remains very small in
comparison with the zero momentum frequency $i\omega_2(0)\approx
2/\tau_s$. In contrast, the gap develops rapidly if we move along
a direction in parameter space corresponding to
$\gamma_R=\gamma_D$ (green line). This observation suggests that,
in order  to observe experimentally the ``absence'' of spin
relaxation at a certain finite momentum, the minimization of the
cubic Dresselhaus interaction should be a major concern.
In addition, we have to remember about the small momenta
relaxation times given by equations (\ref{Dzz0kapp}) and
(\ref{iw0kapp}), in particular the one associated with the
$i\omega_1$ mode. To summarize the results concerning relaxation
times, we refer to the diagram in Fig. \ref{FIGsr1}. For a uniform
system (k=0), the only mode that contributes to the dynamics is
$i\omega_3$, and the corresponding spin relaxation time of the
out-of-plane component $S_z$ is $\tau_s/2$, regardless of the
spin-orbit coupling parameters. The system can have a slower
relaxation at non-zero momenta that can be either arbitrarily
small or of order $1/L_s$, depending on the values of the coupling
parameters $\gamma_R$ and $\gamma_D$. In the red region, all three
modes $i\omega_j({\bf k})$ have the minimum at $k=0$ and the
largest spin relaxation time,
$\tau_{max}(0)=\tau_s/(1-|\gamma_R\gamma_D|/2)$, is obtained at
small k-vectors. In the blue (yellow) region, in addition to this
small momentum minimum, we have the minima at $k_+^0$ ($k_-^0$)
given by equations (\ref{MINk}) and (\ref{iwMINk}) and to
determine the slowest spin relaxation we have to compare
$\tau_{max}(0)$ with
$\tau_{max}(k_{\pm})=\tau_s/i\omega_{2(1)}(k_{\pm}^0)$. Finally,
in the green regions, the largest spin relaxation time is
associated with one of the minima at $k_{\pm}^0$. The diffusion is
isotropic if $\gamma_R\gamma_D=0$ and anisotropic otherwise,
especially in the vicinity of the symmetry points.

\subsection{Real space picture}
\label{rspace}

\begin{figure}
\begin{center}
\includegraphics[width=0.6\textwidth]{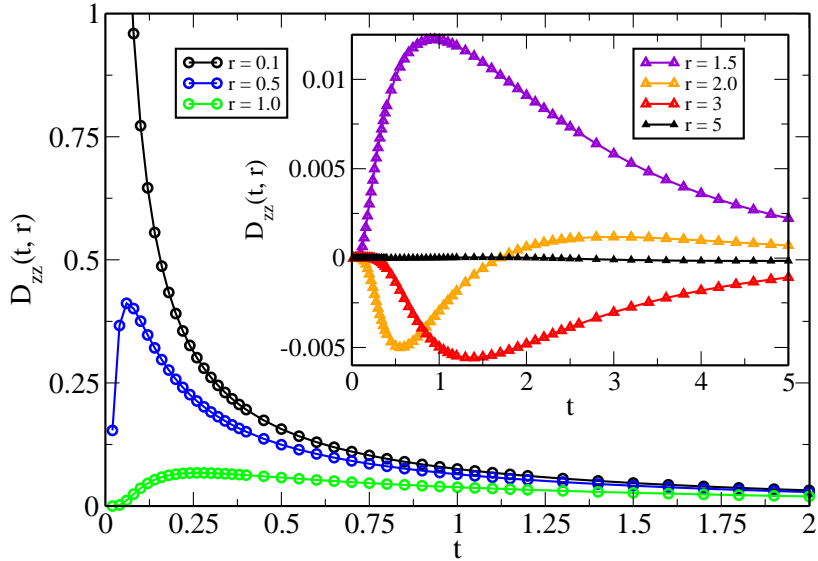}
\caption{Relaxation of the out-of-plane spin density in the pure
Rashba case. The spin density $\rho_z$, having an initial
$\delta$-like profile in real space, is injected in the origin at
$t=0$. The curves show the time evolution $\rho_z(t, r)=D_{zz}(t,
r)$ at various distances r from the origin.  The large time
behavior is described by Eq.~(\ref{Disotrop}) for $\gamma=2$. Time
is expressed in units of $\tau_s$ and length in units of $L_s$.}
\label{FIGsr3}
\end{center}
\end{figure}

\begin{figure}
\begin{center}
\includegraphics[width=0.6\textwidth]{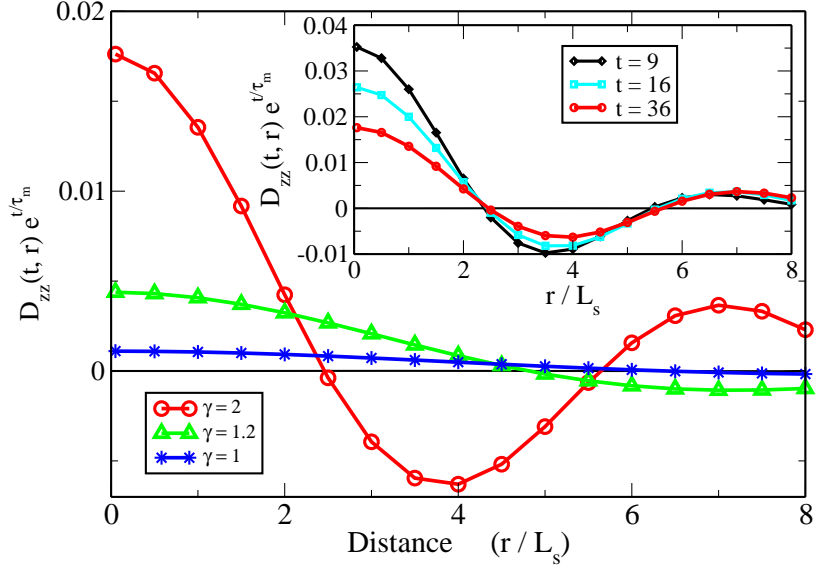}
\caption{Real space dependence  of the out-of-plane spin density
in a system with isotropic spin-orbit interaction. The curves
represent $D_{zz}(t, r)$ multiplied by the exponential factor
$\exp(t/\tau_m)$, where $t=36$ (in units of $\tau_s$) and
$\tau_m=(6\gamma2-\gamma4-1)/(4\gamma2)$ represents the maximum
relaxation time corresponding to the finite $k^0$ minimum of the
relaxation rate. At short distances, $r\ll\sqrt{t}=6$, the
solutions are described by Eq. (\ref{Disotrop}). The period of the
spatial oscillations increases as we decrease $\gamma$: $\gamma=2$
(pure Rashbe or pure linear Dresselhaus interaction) - red line
and circles, $\gamma=1.2$ - green line and triangles, $\gamma=1$ -
blue line and stars. Notice that, as we approach the
non-oscillatory regime characterized by $\gamma<1$, the spin
density becomes independent of distance in the asymptotic limit
$t\gg r^2$ (see Eq. (\ref{Disotrop})) and the $\gamma=1$ curve).
The inset shows the spatial dependence of the spin density of a
pure Rashba system ($\gamma=2$) at different times. Again, we
multiplied the results by the exponential factors
$\exp(t/\tau_m)$. Notice that in the limit described by Eq.
(\ref{Disotrop}) the curves are proportional, with proportionality
factors 1/3 (black curve and diamonds, $t=9$), 1/4 (cyan with
squares, $t=16$) and 1/6 (red line and circles, $t=36$).}
\label{FIGsr4}
\end{center}
\end{figure}

We turn now our attention to the real space behavior of
${D}_{zz}$, corresponding to the diffusion of a delta
function-like spin density injected in the origin at $t=0$.  While
the analytical treatment of the general case its rather
complicated, we can easily solve special cases, like the isotropic
case ($\gamma_R\gamma_D=0$), or the symmetry point
$\gamma_R=\pm\gamma_D=\sqrt{2}$. In particular, for the isotropic
case we obtain from Eq.~(\ref{Dijm1}), after inverting the matrix
and performing a Fourier transform with respect to the frequency,
\beq {D}_{zz}(t, {\bf k}) =
\frac{1}{2}\left[1+\frac{1}{\sqrt{1+4\gamma^2 k^2}}\right]
e^{-\left(\frac32+k^2+\frac12\sqrt{1+4\gamma^2 k^2}\right)t} +
\frac{1}{2}\left[1-\frac{1}{\sqrt{1+4\gamma^2 k^2}}\right]
e^{-\left(\frac32+k^2-\frac12\sqrt{1+4\gamma^2 k^2}\right)t},
\label{Dzzisotr}\eeq where $\gamma=\gamma_R$ if $\gamma_D=0$ and
$\gamma=\gamma_D$ if we have a pure Dresselhaus spin-orbit
coupling. The real space dependence is obtained after performing
another Fourier transform, with respect to momentum, which reduces
to the one dimensional integral \beq {D}_{zz}(t, {\bf r}) =
\frac{1}{4\pi}\int_0^{\infty}dk~k~J_0(kr)\mbox{D}_{zz}(t, {\bf
k}), \eeq where $J_0(kr)$ represents a Bessel function of the
first kind. The time dependence of ${D}_{zz}(t, {\bf r})$ at
distance $r$ form the origin is shown in Fig. \ref{FIGsr3} for the
particular case of pure Rashba coupling $\gamma=2$.
The curves represent the time evolution of the out-of-plane  spin
density at different positions in the plane after spin was
injected in the origin at $t=0$. Notice the oscillations in the
direction of the spin polarization (see Fig.~\ref{FIGsr4}). The
analytical expressions for these oscillations can be derived in
the large time limit using a saddle point approximation for the
integral over momenta  [more precisely in the limit $t/\tau_s \gg
(r/L_s)^2$]. For the isotropic case this asymptotic behavior has
the form \beqa {D}_{zz}(t, {\bf r}) &\approx&
\frac{\gamma^2-1}{8\gamma\sqrt{\pi
t}}J_0\left(\frac{\sqrt{\gamma^4-1}}{2\gamma}~r\right)
e^{-\frac{6\gamma^2-\gamma^4-1}{4\gamma^2}~t},
~~~~~~~\mbox{if}~~~|\gamma|>1, \nonumber \\
{D}_{zz}(t, {\bf r}) &\approx&
\frac{\gamma^2}{4\pi(1-\gamma^2)t^2}~e^{-t},
~~~~~~~~~~~~~~~~~~~~~~~~~~~~~~~~~~\mbox{if}~~~|\gamma|<1,
\label{Disotrop} \eeqa  where $\gamma=\gamma_R$ or
$\gamma=\gamma_D$, as appropriate. The asymptotic behavior is
qualitatively different for $|\gamma_R|>1$ (or $|\gamma_D|>1$),
the green regions in Fig. \ref{FIGsr1}, and $|\gamma_R|<1$ (or
$|\gamma_D|<1$), red zone. In the first case, ${D}_{zz}(t, {\bf
r})$ oscillates in space at large distances away from the origin,
$r/L_s\gg 1$, with a period $\lambda = 2\pi/(L_s k^0)$ and an
amplitude that decays as $\exp[-i\omega(k^0)t]/\sqrt{r t}$, where
$k^0=k_+^0=k_-^0$ and
$i\omega(k^0)=i\omega_1(k_-^0)=i\omega_2(k_+^0)$ are the
parameters for the finite momentum minima of the dispersion
relations described by equations (\ref{MINk}) and (\ref{iwMINk}).
By contrast, in the red region, spin diffusion at large times,
$t/\tau_s \gg (r/L_s)^2$, is independent of position. In addition,
the spin relaxation time has the value $\tau_s$, independent of
the spin-orbit coupling, and the time dependence of the pre-factor
is $1/t^2$, instead of the $1/\sqrt{t}$ dependence for the
oscillating case.

\begin{figure}
\begin{center}
\includegraphics[width=0.4\textwidth]{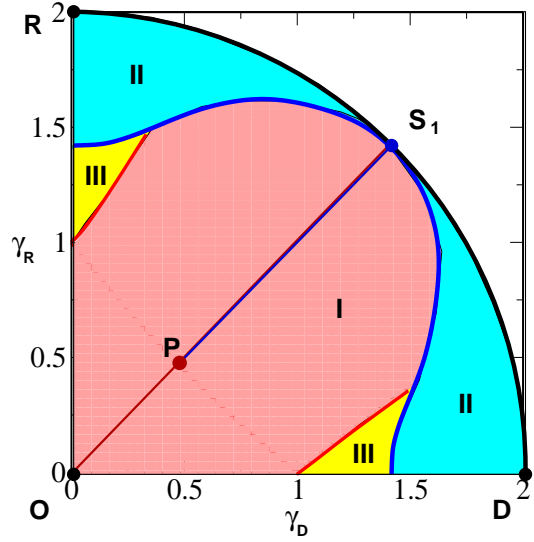}
\caption{Diagram summarizing the asymptotic relaxation of the
out-of-plane spin density in real space for a quarter of the
parameter space ($\gamma_R\ge0, ~\gamma_D\ge0$). In the pink zone
(except the segment $OS_1$ corresponding to $\gamma_R=\gamma_D$)
the solution is anisotropic at short times and becomes isotropic
at large times. This solution is controlled by the zero momentum
minimum $i\omega_1(0)=1-\gamma_R\gamma_D/2$ and has the asymptotic
expression given by Eq.~(\ref{redasymm}). For the segment $OP$,
the solution is controlled by the zero momentum minimum
$i\omega_2(0)=1+\gamma_R\gamma_D/2$ and has the same general
characteristics as the neighboring pink zone. In the blue region
the solution is anisotropic at all timescales and features real
space oscillations along the '+' direction. The long time dynamics
is controlled by the finite momentum minimum $i\omega_2(k_+^0)$
and the asymptotic behavior is described by Eq.~(\ref{Dtr2k}). The
same type of solution is valid for the $PS_1$ segment. Finally, in
the yellow zone, the solution has real space oscillations along
the '-' direction and an asymptotic behavior controlled by the
finite momentum minimum $i\omega_1(k_-^0)$. Notice that, as we
approach the axes $\gamma_R=0$ and $\gamma_D=0$ the spin
relaxation, as described by Eq.~(\ref{Dzzisotr}),  becomes
isotropic. We stress that the regimes represented here are well
defined in the long time limit. At some intermediate timescales,
and especially near the zone boundaries, the solution may exhibit
mixed characteristics when the minima of two distinct relaxation
rate modes have comparable contributions.} \label{FIGsr5}
\end{center}
\end{figure}

For $\gamma_R\gamma_D\neq 0$ the dynamics of the $S_z$ spin
component is anisotropic. The simplest analytical treatment of
such a case obtains at the symmetry points
$|\gamma_{R(D)}|=\sqrt{2}$. Starting with Eq.~(\ref{calDr}), we
get, after inverting the matrix and performing a Fourier transform
with respect to the frequency, \beq {D}_{zz}(t,{\bf k})=\frac12
e^{-\left[(\sqrt{2}+k_+)^2+k_-^2\right]t}+ \frac12
e^{-\left[(\sqrt{2}-k_+)^2+k_-^2\right]t}, \label{Dzztksym} \eeq
where we assumed that $\gamma_R\gamma_D=2$, which corresponds to
the symmetry points $S_1$ and $S_1^{\prime}$ in Fig. \ref{FIGsr1}.
A similar equation can be obtained for $\gamma_R\gamma_D=-2$ by
exchanging $k_+$ and $k_-$. This equation does not acquire any
spin-charge coupling corrections, as the coupling vanishes
 at the symmetry points, $\mu_0=\mu_1=0$, regardless of the
 strength of the spin-orbit interaction.
Next,  after performing the Fourier transform with respect to
momentum, we obtain the exact real space dependence at the
symmetry point, \beq {D}_{zz}(t, {\bf r}) = \frac{1}{4\pi
t}~e^{-\frac{r^2}{4t}}\cos(\sqrt{2}~r_+),  \eeq where
$r^2=r_+^2+r_-^2$ and we assumed that $\gamma_R\gamma_D=2$.
Because of the gapless minimum of the $i\omega_2(k_+)$ mode, the
large time decay is proportional to $1/t$, rather than
exponential. In addition, the spin density oscillates along the
$r_+$ direction, due to  the finite momentum minimum, while it
becomes independent of $r_-$ in the large time limit. For a
general anisotropic case, the asymptotic behavior is determined by
the minimum of the lowest relaxation rate. A characterization of
the locations of the lowest minimum in the plane of the spin-orbit
coupling parameters is  provided in Fig. \ref{FIGsr5}. We
determine the asymptotic spin dynamics using a quadratic
approximation for the dispersion curves in the vicinity of the
minima. The approximate solutions of the secular equation
$\Delta(s, {\bf k})$ for wave-vectors with arbitrary $k_+$ and
small $k_-$ components are given in Appendix \ref{AppA}. Assuming
that $\gamma_R\gamma_D>0$, the zero momentum minimum of the
$i\omega_1$ mode represents the lowest energy contribution for a
system with spin-orbit coupling parameters corresponding to the
pink zone in Fig. \ref{FIGsr5}. The asymptotic $S_z$-spin
diffusion in the zone~I of the parameter space (see
Fig.~\ref{FIGsr5}) is described by \beq {D}_{zz}^I(t, {\bf r})
\approx \frac{2tC_0^{(1)}-r_-^2}{16\pi
t^3}~\frac{(\gamma_R-\gamma_D)^2}
{\left(1+\frac{\gamma_R\gamma_D}{2}\right)^2(C_0^{(1)})^{\frac{5}{2}}}~
e^{-\frac{1}{4t}\left(r_+^2+\frac{r_-^2}{C_0^{(1)}}\right)-
\left(1-\frac{\gamma_R\gamma_D}{2}\right)t}, \label{redasymm}\eeq
where the coefficient $C_0^{(1)}$ is given by Eq.~(\ref{appC10}).
Eq.~(\ref{redasymm}) is valid for times larger than the spin
relaxation time, $t\gg\tau_s$, when $\gamma_R\neq\gamma_D$. When
the couplings are equal, the $i\omega_1$ mode whose $k=0$ minimum
generates the asymptotic behavior described by
Eq.~(\ref{redasymm}) does not contribute to the spin dynamics.
Instead, for $\gamma_R=\gamma_D$ corresponding to the $OP$ segment
in Fig. \ref{FIGsr5}, the lowest minimum is the zero momentum
minimum of the $i\omega_2$ mode and it will generate an expression
similar to Eq.~(\ref{redasymm}), but with $r_{\pm}\rightarrow
r_{\mp}$ and $\gamma_D\rightarrow-\gamma_D$. Notice that in this
regime spin diffusion is anisotropic only for times of order $r^2$
or smaller and becomes isotropic at longer times. Finally, for the
anisotropic regime controlled by a finite momentum minimum in the
dispersion relations, the blue and yellow  regions and the segment
$PS_1$ in Fig. \ref{FIGsr5}, the asymptotic expression for
${D}_{zz}$ can be obtained using the quadratic expansions given in
Appendix \ref{AppA}, in particular equations (\ref{appw2k}) and
(\ref{appB2C2}). We obtain the asymptotic expression for the
diffuson in the regime described by zone~II of the parameter space
(see Fig.~\ref{FIGsr5}) \beq \mbox{D}_{zz}^{II}(t, {\bf r})\approx
\frac{1}{4\pi t}\frac{D_{k_+^0}^{(2)}}
{\sqrt{B_{k_+^0}^{(2)}~C_{k_+^0}^{(2)}}}
\exp\left[-\frac{1}{4t}\left(\frac{r_+^2}{B_{k_+^0}^{(2)}}+\frac{r_-^2}{C_{k_+^0}^{(2)}}\right)
-i\omega_2(k_+^0)t\right] \cos(k_+^0r_+), \label{Dtr2k} \eeq where
$D_{k_+^0}^{(2)}=1-(1-\gamma_R\gamma_D/2)/(\gamma_R+\gamma_D)^2$,
$B_{k_+^0}^{(2)}$ and $C_{k_+^0}^{(2)}$ are given in Appendix
\ref{AppA}, and $i\omega_2(k_+^0)$ represents the $k_+^0$ minimum
of the $i\omega_2$ mode given explicitly by equations (\ref{MINk})
and (\ref{iwMINk}). Equation (\ref{Dtr2k}) describes the large
time dynamics of the out-of-plane spin component when the
$i\omega_2(k_+^0)$ represents lowest energy minimum, i.e. for the
blue zone and the segment $PS_1$ in Fig. \ref{FIGsr5}. A similar
expression can be obtained when the dynamics is controlled by the
$i\omega_1(k_-^0)$ minimum (yellow zone~III) using the
correspondence rules $i\omega_2\rightarrow i\omega_1$, $r_+
\leftrightarrow r_-$ and $\gamma_D\rightarrow -\gamma_D$: \beq
\mbox{D}_{zz}^{III}(t, {\bf r})\approx \frac{1}{4\pi
t}\frac{D_{k_-^0}^{(2)}} {\sqrt{B_{k_-^0}^{(2)}~C_{k_-^0}^{(2)}}}
\exp\left[-\frac{1}{4t}\left(\frac{r_-^2}{B_{k_-^0}^{(2)}}+\frac{r_+^2}{C_{k_-^0}^{(2)}}\right)
-i\omega_1(k_-^0)t\right] \cos(k_-^0r_-), \label{Dtryellow} \eeq
where
$D_{k_-^0}^{(2)}=1-(1+\gamma_R\gamma_D/2)/(\gamma_R-\gamma_D)^2$
and the coefficients $B_{k_-^0}^{(2)}$ and $C_{k_-^0}^{(2)}$ are
given in Appendix \ref{AppA}.
 Notice that Eq.~(\ref{Dtr2k},\ref{Dtryellow})
reduce to Eq.~(\ref{Dzztksym}) at the symmetry points
$\gamma_R=\gamma_D=\pm\sqrt{2}$ where it becomes exact. Within
this regime, the solution of the diffusion equation is
characterized by real space oscillations with a period
$\lambda=2\pi L_s/k_+^0$ determined by the position of the minimum
in k-space while,  at times much larger than $\tau_s ~r^2/L_s^2$,
the solution becomes independent of $r_-$. The relaxation time,
$\tau_s/(i\omega_2(k_+^0))$, can be much larger than the
relaxation time in the regime controlled by the zero momentum
minimum, $\tau_s/(i\omega_1(0))$, and diverges as we approach the
symmetry points. On the other hand, when we approach the
$\gamma_R=0$ or $\gamma_D=0$ axes, the solution becomes isotropic
and some of the approximations used in deriving equations
(\ref{redasymm}) and (\ref{Dtr2k}) are not longer valid. Instead,
the asymptotic behavior in the isotropic regime is described by
Eq.~(\ref{Disotrop}).

\section{The role of the spin-charge coupling}\label{SecIV}

In this section we study the effects of spin-charge coupling on
spin and charge diffusion. The main results of this section are
equations (\ref{s14}) and (\ref{iwk14}) describing the behavior of
the relaxation rate modes in the presence of the spin-charge
coupling.

\begin{figure}
\begin{center}
\includegraphics[width=0.6\textwidth]{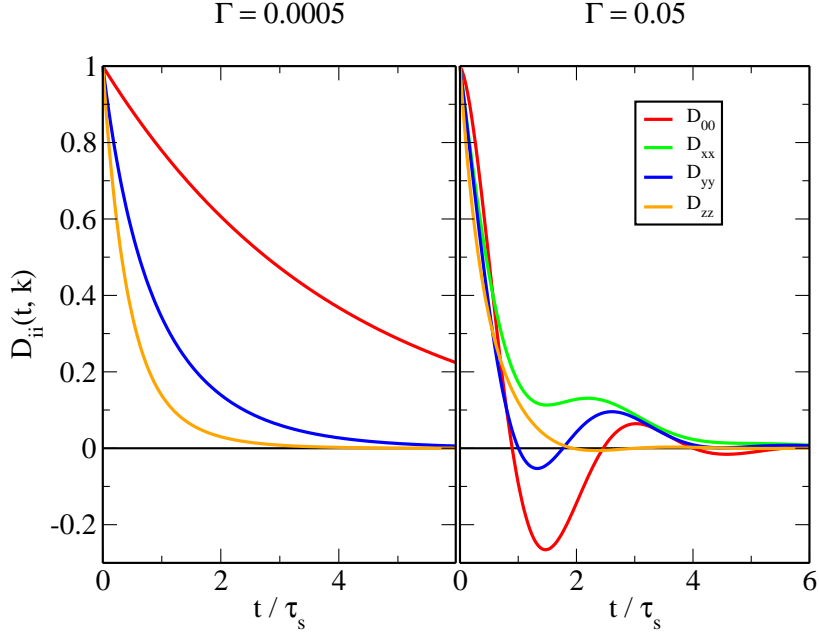}
\caption{Time dependence of the diagonal components of the
diffusion matrix in the weak spin-orbit coupling (left panel) and
strong spin-orbit coupling (right panel) regimes for
$\gamma_R=1=-\gamma_D$. The matrix elements are evaluated at a
momentum ($k_x$, $k_y$) = ($0.5/L_s$, 0). In the weak spin-orbit
coupling limit all the quantities decay exponentially, while in
the opposite  limit  they exhibit oscillatory behavior. Notice
that due to the particular choice of parameters, in the
weak-coupling limit the in-plane spin components have identical
relaxation curves, $D_{xx}=D_{yy}$, while in the strong-coupling
limit the two quantities differ due to the k-dependent coupling to
the charge channel.} \label{FIGsr7}
\end{center}
\end{figure}

In the absence of spin-charge coupling, the charge diffuses
according to the standard diffusion equation, while a spin density
characterized by a given k-vector decays, in general, as a sum of
three exponentials. Recent spin grating
experiments\cite{WeberAwschal} have shown that, in certain
conditions, the time decay of an optically injected spin density
may include an oscillatory component. In this section, we explore
the possibility that the appearance of these oscillations is due
to spin-charge coupled dynamics. To this end, we re-derive the
Green's function of the diffusion equation starting from the
polarizability matrix Eq.~(\ref{Dijm1}), or from
Eq.~(\ref{calDr}), but without neglecting the spin-charge coupling
terms proportional to $\mu_0$ and $\mu_1$. Inverting the
polarizability matrix produces, in general, a complicated
expression for the diffuson. However, there are several special
cases in which significant simplifications occur. These cases
involve, on the one hand, special values of the spin-orbit
coupling parameters, namely the pure Rashba interaction,
$\gamma_D=0$, $\gamma_R=\pm 2$, the pure Dresselhaus coupling,
$\gamma_R=0$, and the symmetry points $\gamma_R = \pm \sqrt{2}$,
$\gamma_D = \pm \sqrt{2}$. On the other hand, the inversion of
$\Pi(\omega, {\bf k})$ can can be performed analytically for {\it
arbitrary} values of the spin-orbit coupling parameters along the
special directions in momentum space corresponding to $k_+$ and
$k_-$. Our strategy is to obtain explicit analytical expressions
for these special cases, and use them, in conjunction with
numerical results for arbitrary sets of parameters, to extract and
characterize the main features of the general solution.
 We start with  the isotropic case corresponding to a pure Rashba
 coupling, or a  pure Dresselhaus spin-orbit interaction.
The modes that control the coupled spin-charge dynamics are the
solutions of the secular equation \beq \Delta(s, {\bf k}) \equiv
\left[s(s+1)-\gamma^2k^2\right]\left[s(s-1)+g^2\mu_l^2k^2\right]=0,
\label{DelskR} \eeq where $\Delta(s, {\bf k})$ is the determinant
of the $\hat{\Pi}(s, {\bf k})$ matrix, $s=-i\omega+1+k^2$ and
$\gamma=\pm 2$ for the Rashba case, or $\gamma=\gamma_D\in[-2,2]$
for a pure Dresselhaus spin-orbit interaction. The spin-charge
coupling is measured by the parameter $\mu_l$ and we have
$\mu_l=\mu_1=\mbox{sign}(\alpha)\alpha^2$ in the Rashba case and
$\mu_l=\mu_0=\left[(3\beta_3-\beta_1)(\beta_3^2-\beta_1^2)-\beta_1\beta_3^2\right]
/\left[(\beta_1-\beta_3)^2+\beta_3^2\right]^{1/2}$ for the
Dresselhaus coupling.
\begin{figure}
\begin{center}
\includegraphics[width=0.6\textwidth]{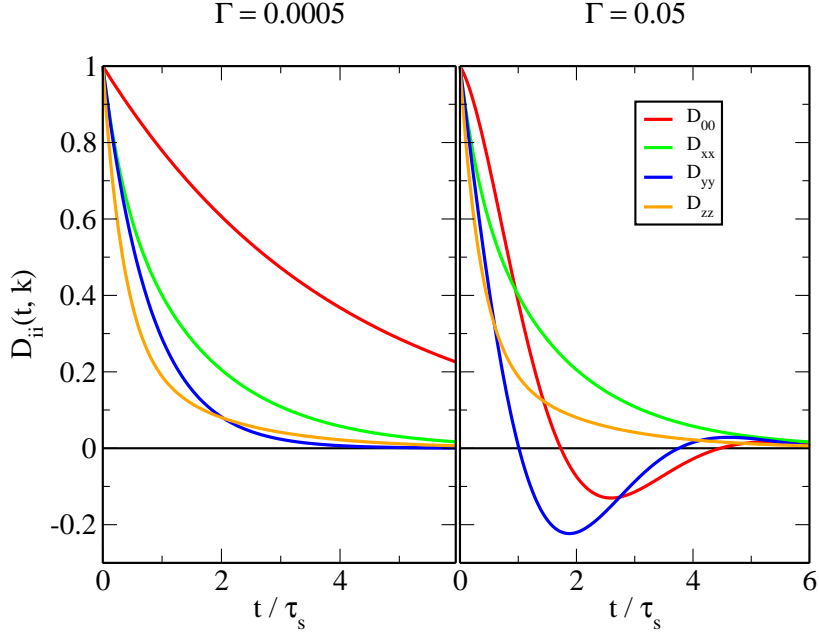}
\caption{Time dependence of the diagonal components of the
diffusion matrix in the weak spin-orbit coupling (left panel) and
strong spin-orbit coupling (right panel) regimes. The matrix
elements are evaluated for the pure Rashba case ($\gamma_R=2$,
$\gamma_D=0$) at a momentum ($k_x$, $k_y$) = ($0.5/L_s$, 0). In
the weak spin-orbit coupling limit all the quantities decay
exponentially, while in the opposite  limit the charge and the
$S_y$ channels exhibit oscillatory behavior. In the pure Rashba
(or pure Dresselhaus) case the out-of-plane spin component does
not couple to the charge, due to symmetry. For ${\bf k}$ parallel
to the x-axis, this is also the case for the in-plane $S_x$ spin
component (see Eq.~(\ref{AppDij})).} \label{FIGsr6}
\end{center}
\end{figure}
The full expression of the diffusion matrix is given in Appendix
\ref{AppB}. The real time behavior of the diffusion propagator can
be obtained after performing a Fourier transform of (\ref{AppDij})
with respect to the frequency. This time dependence is determined
by the roots $s_j=-i\omega_j+1+k^2$ of equation (\ref{DelskR}),
and has the general form (\ref{genSol}). For the pure cases
described by the equations (\ref{DelskR}) and
 (\ref{AppDij}), the four modes that control the coupled spin-charge
 dynamics are \beq \left\{\begin{array}{lll}
i\omega_0({\bf k}) =\frac{1}{2}+k^2-\frac{1}{2}\sqrt{1-4g^2\mu_l^2 k^2} \\ ~~\\
i\omega_1({\bf k}) =\frac{1}{2}+k^2+\frac{1}{2}\sqrt{1-4g^2\mu_l^2 k^2} \\ ~~\\
i\omega_2({\bf k}) =\frac{3}{2}+k^2-\frac{1}{2}\sqrt{1+4\gamma^2 k^2} \\~~\\
i\omega_3({\bf k}) =\frac{3}{2}+k^2+\frac{1}{2}\sqrt{1+4\gamma^2
k^2}\end{array} \right. \label{s14} \eeq where $g=2v_Fk_F\tau$,
$k^2=k_x^2+k_y^2$ and the k-vector is expressed, as usual,  in
units of $1/L_s$. Equation (\ref{s14}) has several key features.
First, we notice that the mode responsible for the charge dynamics
in the absence of  spin-charge coupling, namely $i\omega_0$,
couples with only one spin mode, $i\omega_1$, while the other spin
modes, $i\omega_2$ and $i\omega_3$ remain unaffected by the
spin-charge coupling. Consequently, the modes  $i\omega_2$ and
$i\omega_3$ are real and positive for any value of the momentum,
and the corresponding terms in Eq.~(\ref{genSol}) will decay
exponentially in time. On the other hand, the solutions
$i\omega_0$ and $i\omega_1$ are real only if  $2g\alpha^2 k_F L_s
= 2\tau v_F k_F\alpha < 1$. This condition is always satisfied
 in the weak spin-orbit coupling regime, in which the diffusion
 equation formalism applies. We conclude that within  the diffusive
 regime a system will never exhibit oscillatory dynamics. Below, we show
 that this conclusion holds for arbitrary spin-orbit coupling parameters.
Nonetheless, we can formally extrapolate Eq.~(\ref{s14}) outside
its strict domain of validity and apply it to the strong
spin-orbit interaction regime. In this limit, i.e. for $\tau v_F
k_F\alpha > 1$, the modes $i\omega_0$ and $i\omega_1$ acquire an
imaginary component, which generates oscillatory terms in the
solution (\ref{genSol}). We interpret this second major feature of
Eq.~(\ref{s14}) as an indication that oscillatory dynamics is a
signature of strong spin orbit interactions. While the diffusion
equation formalism cannot offer a quantitative description of this
 regime, it should represent a good indicator for the range of
 parameters where the oscillatory behavior is likely to appear and
 for the degree in which various components of the spin and charge
 sectors are affected. For example, from Eq.~(\ref{AppDij}) we
 observe that the time evolution of the diagonal $S_z$ component $D_{zz}$
is controlled by the modes $i\omega_2$ and $i\omega_3$, which do
not couple with the charge. Consequently, in the pure Rashba (or
pure Dresselhaus) case the out-of-plane spin component does not
exhibit oscillatory dynamics for any strength of the spin-orbit
coupling. This result is connected to the isotropic nature of the
spin-orbit splitting in these particular cases. In contrast, the
dynamics of the in-plane spin components is coupled with the
charge dynamics. Explicitly, the time dependence of the matrix
element ${D}_{00}$ describing charge relaxation is given by \beq
{D}_{00}(t, {\bf k}) = \left\{\begin{array}{ll}
\left[\cosh\left(\frac{t}{2}\sqrt{1-4g^2\mu_l^2 k^2}\right)
+\frac{\sinh\left(\frac{t}{2}\sqrt{1-4g^2\mu_l^2
k^2}\right)}{\sqrt{1-4g^2\mu_l^2
k^2}}\right]e^{-\left(\frac12+k^2\right)t},
~~~~~~~~~~~~~~~~\mbox{if}~~~~~~~~ 2g\mu_l k < 1, \\ ~~\\
\left[\cos\left(\frac{t}{2}\sqrt{4g^2\mu_l^2 k^2-1}\right)
~+~\frac{\sin\left(\frac{t}{2}\sqrt{4g^2\mu_l^2
k^2-1}\right)}{\sqrt{4g^2\mu_l^2
k^2-1}}\right]e^{-\left(\frac12+k^2\right)t},
~~~~~~~~~~~~~~~~\mbox{if}~~~~~~~~ 2g\mu_l k > 1,
\end{array} \right. \label{D11R} \eeq
where the wave vector k is expressed in units of $1/L_s$ and time
in units of spin relaxation time, $\tau_s$. Notice that in the
absence of spin-charge coupling, ${D}_{00}$ reduces to the
standard solution of the diffusion equation, ${D}_{00}(t, {\bf k})
= \exp(-k^2 t)$, while the time dependence of the matrix elements
associated with the spin degrees of freedom is determined by $s_2$
and $s_3$ only. This behavior is illustrated in Fig. \ref{FIGsr6}
which shows the time dependence of all diagonal matrix elements
$\hat{D}_{ii}(t, {\bf k})$ for ${\bf k} = (0.5, 0)$. The curves in
the left panel were calculated in the weak spin-orbit coupling
regime corresponding to $\Gamma=0.0005$ and $g=1000$. Notice that
in the charge channel we have the standard exponential decay,
$\exp(-k^2 t)$, while in the spin channels the effect of the
finite relaxation time $\tau_x=\tau_y=\tau_z/2$ is evident. By
contrast, in the strong-coupling limit, the charge and  $S_y$
channels exhibit oscillatory behavior. Also notice that, as a
result of the spin-charge coupling, the charge relaxes on the same
time scale as the spin degrees of freedom. We mentioned above that
for a pure Rashba (or pure Desselhaus) spin-orbit coupling, the
charge does not couple with the out-of-plane spin component due to
symmetry. In the example shown in Fig. \ref{FIGsr6} similar
symmetry reasons apply to the $S_x$ spin component, due to the
particular choice of momentum along the x-axis. However, in
general all spin components  will  couple to the charge. For
example, the general expression for the ${D}_{0z}$ matrix element
describing the coupling between  charge and out-of-plane spin
density is \beq {D}_{0z}(s, {\bf k})=\frac{g}{2\Delta(s,{\bf
k})}(k_x^2-k_y^2)[\gamma_R\gamma_D(\gamma_D\mu_0-\gamma_R\mu_1)+2s(\gamma_D\mu_1-\gamma_R\mu_0)],\eeq
where $\Delta(s,{\bf k})$ is the determinant of $\hat{D}^{-1}$.
Let us consider a set of parameters for which ${D}_{0z}$ is
non-zero. We show in Fig. \ref{FIGsr7} the time dependence of the
diagonal matrix elements in the presence of both Rashbe and
Dessselhaus terms, so that $\gamma_R=1$ and $\gamma_D=-1$. In
terms of the original coupling constants, this case corresponds to
$\alpha\approx0.47~\Gamma$, $\beta_1\approx0.19~\Gamma$, and
$\beta_3\approx0.66~\Gamma$.
Notice that in the strong-coupling limit (right panel in Fig.
\ref{FIGsr7}) all channels exhibit oscillatory behavior, although
the relative amplitude for the $S_z$ component is very small.
Again, we observe that ${D}_{00}$ decays on a time scale of order
$\tau_s$, instead of the standard exponential decay $\exp(-k^2
t)$. We conclude that strong spin-orbit interaction can generate
two effects: 1) an oscillatory structure in the decay curves, and
2) an increase in the decay rate of the charge channel. The first
effect should be observable above a certain critical momentum of
order $k_c \approx g\mu_i/L_s$. The second effect can be
understood as a re-normalization of the diffusion constant. In the
limit of small momenta equation (\ref{D11R}) reduces to
$$
{D}_{00}(t, {\bf k}) \approx \exp[-(1+2g^2\mu_l^2)k^2t].
$$
Consequently, in this limit the charge diffuses at long
lengthscales according to the standard diffusion equation but with
a re-normalized diffusion constant. This effective diffusion
constant is enhanced by a factor $1+2g^2\mu_l^2$ over its bare
value:
\begin{equation}
\label{Deff} {\cal D}_c^{\rm eff} \approx
\left(1+2g^2\mu_l^2\right) \frac{v_F^2 \tau}{2}. \end{equation}
Formally, the enhancement factor can be very large. However, we
have to remember that these conclusions are based on an
extrapolation of the diffusion equation formalism outside its
domain of validity and, therefore, should be seen as having
qualitative rather than quantitative relevance. Strictly speaking,
in the strong spin-orbit coupling regime the diffusion equation
formalism breaks down and the notion of a diffusion coefficient
becomes meaningless. The coupled spin-orbit dynamics is no longer
described by a set of differential equations of type (\ref{Dijm1})
and (\ref{calDr}).

Our calculation using the diffusion equation formalism indicates
that oscillatory dynamics may appear in a system with strong
spin-orbit interaction as a consequence of spin-charge coupling.
The effective strength of this coupling depends not only on the
overall strength of the spin-orbit interaction, i.e. on $\Gamma$,
but also on the interplay between different types of spin-orbit
couplings. The net result of these competing effects is contained
in the expressions of the spin-charge coupling coefficients
$\mu_0$ and $\mu_1$ given by Eq.~(\ref{mus}). In the examples
considered above (see Fig. \ref{FIGsr6} and \ref{FIGsr7}), we have
$(\mu_0, ~\mu_1) =(0, ~\Gamma^2)$ for the pure Rashba case and
$(\mu_0, \mu_1) =(1.09~\Gamma^2, ~1.40~\Gamma^2)$ for the
$\gamma_R=1=-\gamma_D$ case. For comparison, the maximum values of
the spin-charge coupling coefficients are $|\mu_0|\approx
1.10~\Gamma^2$ and $|\mu_1|\approx 1.65~\Gamma^2$. However, for
other values of the spin-orbit coupling parameters these
coefficients, and implicitly the spin-charge coupling effects, may
be much weaker. In particular the special case corresponding to
the symmetric points defined by $\gamma_R=\gamma_D=\pm\sqrt{2}$
deserves special attention (similar considerations can be made for
the case $\gamma_R=-\gamma_D$ by interchanging the '+' and '-'
components). In this case $\mu_0=\mu_1=0$ and, consequently,  the
spin-charge coupling vanishes. Using  the rotated coordinate
system and
 expression (\ref{calDr}) for the polarizability matrix, we obtain
 for the diffuson \beq
\hat{D}^{(r)}(\omega, {\bf k}) =\left(
\begin{array}{cccc}
  \frac{1}{s-1} & 0 & 0 & 0 \\
  0 & \frac{s+1}{(s+1)^2-8k_+^2} & 0 & \frac{2\sqrt{2}ik_+}{(s+1)^2-8k_+^2} \\
  0 & 0 & \frac{1}{s-1} & 0 \\
  0 & \frac{-2\sqrt{2}ik_+}{(s+1)^2-8k_+^2} & 0 & \frac{s+1}{(s+1)^2-8k_+^2} \\
\end{array}
\right)\eeq The effects of the spin-charge coupling are totally
absent at the symmetry points, regardless of the strength of the
spin-orbit interaction. In addition, the diffusion matrix is
independent of the $k_-$ wave-vector component and has a gapless
mode for $k_+=\pm\sqrt{2}$, as discussed in the previous section.

To understand the effects of spin-orbit coupling in a system with
arbitrary spin-orbit interaction, we calculate the diffusion
matrix corresponding to  wave-vectors along the $k_{\pm}$
directions. From Eq.~(\ref{calDr}) we obtain the general
expression for the matrix $\hat{D}^{(r)}(s, k_{\pm})$ given in
Appendix \ref{AppB}. The real time behavior can be obtained by
Fourier transforming (\ref{AppDwkplus}) with respect to the
frequency. Each matrix element has a time dependence of the form
given by Eq.~(\ref{genSol}), except that this time the k-vector is
restricted to the $k_{\pm}$ directions. The four modes that
control the coupled spin-charge dynamics are \beq
\left\{\begin{array}{lll}
i\omega_0(k_+) =\frac{1}{2}+k_+^2-\frac{\gamma_R\gamma_D}{4}-\frac{1}{2}\sqrt{(1-\frac{\gamma_R\gamma_D}{2})^2-4g^2(\mu_0+\mu_1)^2 k_+^2} \\ ~~\\
i\omega_1(k_+) =\frac{1}{2}+k_+^2-\frac{\gamma_R\gamma_D}{4}+\frac{1}{2}\sqrt{(1-\frac{\gamma_R\gamma_D}{2})^2-4g^2(\mu_0+\mu_1)^2 k_+^2}\\ ~~\\
i\omega_2(k_+)=\frac{3}{2}+k_+^2+\frac{\gamma_R\gamma_D}{4}-\frac{1}{2}\sqrt{(1-\frac{\gamma_R\gamma_D}{2})^2+4(\gamma_R+\gamma_D)^2 k_+^2} \\~~\\
i\omega_3(k_+)=\frac{3}{2}+k_+^2+\frac{\gamma_R\gamma_D}{4}+\frac{1}{2}\sqrt{(1-\frac{\gamma_R\gamma_D}{2})^2+4(\gamma_R+\gamma_D)^2
k_+^2} \end{array} \right. \label{iwk14} \eeq where $\mu_0$ and
$\mu_1$ are expressed in terms of the spin-orbit coupling
parameters in Eq.~(\ref{mus}). Similar expressions can be written
for the $k_-$ direction, using the correspondence rules
'+'$\rightarrow$'-', $\gamma_D\rightarrow -\gamma_D$,
$\mu_1\rightarrow -\mu_1$ and $i\omega_1\leftrightarrow
i\omega_2$. As in the isotropic case discussed above, the charge
mode $i\omega_0$ couples with only one spin mode, $i\omega_1$ for
the $k_+$ direction and $i\omega_2$ for the $k_-$ direction, while
the other spin modes remain unaffected by the spin-charge
coupling. Also, for these special k-space directions, the dynamics
of the out-of-plane component of the spin is determined by the
spin modes that do not couple to the charge, i.e. $\mbox{D}_{zz}$
in Eq.~(\ref{AppDwkplus}) depends on $i\omega_2(k_+)$ and
$i\omega_3(k_+)$ only. We conclude that oscillatory dynamics, the
re-normalization of the diffusion constants, or any other
spin-charge coupling effect should not be observable in the
relaxation of an injected out-of-plane spin density with a
wave-vector along the $k_{\pm}$ directions. However, for an
arbitrary direction, the effects of the spin-charge coupling are
present, as we have seen in Fig. \ref{FIGsr7}, as all the modes
contribute to the spin dynamics. The set of four relaxation rate
modes, has some main qualitative features that are independent of
direction: i) there are always at least two purely real modes, one
of them being $i\omega_3$, and ii) an imaginary component may be
acquired by the charge mode $i\omega_0$ and one of the spin modes
$i\omega_1$ or $i\omega_2$, exactly which one of them depending on
the direction in k-space. The appearance of the imaginary
component takes place above a certain minimum value of the
wave-vector satisfying the condition $2g|\mu_0\pm\mu_1|k_{min} =
1\mp\gamma_R\gamma_D/2$. We mention that the existence of such an
imaginary component does not necessarily generates an oscillatory
relaxation of the spin and charge densities. The general solution
in this case contains two exponentially decaying terms an addition
to the oscillatory term with exponentially decaying amplitude and
the existence of an oscillatory component in the relaxation curves
is determined by the relative intensity of these contributions.
Finally, at small values of the wave-vector, the coupled
spin-charge relaxation is always monotonic. However, the
relaxation is, in general, anisotropic and the diffusion
coefficients get re-normalized.

\section{Summary}\label{SecV}

In this paper, we  studied spin diffusion in a generic spin-orbit
coupled system. We  found that there are a number of different
dynamic regimes and phenomena controlled by the relative values of
the spin-orbit coupling parameters. These unusual phenomena
include an enhancement of the spin polarization lifetime at finite
wave-vectors, anisotropic spin diffusion, an oscillatory behavior
of the spin relaxation in real space, an enhancement of the
effective charge diffusion coefficient, and possible real-time
oscillations in spin diffusion dynamics. The existence and
manifestations of these phenomena depend in a non-trivial manner
on the strength of the various terms contributing to the
spin-orbit interaction.

One of the main results of the paper is a general analytic form of
the spin-charge coupled diffusion equation in the presence of all
possible types of spin-orbit couplings. The parameters of this
equation can be conveniently  parameterized by three dimensionless
coupling constants: The overall strength of the spin-orbit
coupling ($\Gamma$) and two  Rashba-type ($\gamma_R$) and
Dresselhaus-type ($\gamma_D)$ contributions to spin precession.
The last two parameters must satisfy the physical constraint
$\gamma_R^2+\gamma_D^2\leq4$. We showed that spin diffusion can be
completely characterized by the pair of these dimensionless
parameters ($\gamma_R$, $\gamma_D$).  We have identified the
domains in the parameter space characterized by either non-zero or
zero k-vector maxima of the spin relaxation spectrum, which
determine the asymptotic dynamics at large times. One particularly
important example we considered is the vicinity of the symmetry
point $|\gamma_R|=|\gamma_D|=\sqrt{2}$, where the lifetime becomes
arbitrarily long for certain finite k-vectors. Our analysis showed
that, in order to observe experimentally this slow spin
relaxation, special attention should be paid to minimize the cubic
Dresselhaus spin-orbit interaction.

We have also qualitatively considered the regime of strong
spin-charge coupling and extrapolated the results of the diffusion
equation formalism to the strong spin-orbit interaction regime. We
found that the enhancement of the charge diffusion coefficient and
the oscillatory spin relaxation observed in recent spin-grating
experiments can be qualitatively understood as a spin-charge
coupling effect. We showed that the coupling of spin and charge is
in general strongly k-dependent and can be characterized by two
parameters $\mu_0$ and $\mu_1$, which depend on the spin-orbit
interaction. These coupling parameters depend not only on the
overall interaction strength $\Gamma$, but also on the interplay
between the Rashba and Dresselhaus contributions. We argued that a
necessary condition to observe an oscillatory behavior is a strong
spin-orbit interaction, corresponding to the crossover from
diffusive transport to the ballistic regime. For a given direction
in k-space, the effect should be observable for k larger than a
certain minimum wave-vector. However, we predict that the effect
is not observable along certain directions in k-space for which
spin-charge coupling is prevented by symmetry. For example, the
out-of-plane spin polarization does not couple with the charge
sector in the pure Rashba or  pure Dresselhaus case. The coupling
is also absent in general for directions in k-space corresponding
to $k_+$ and $k_-$ (see text). In addition, we predict that the
effect is absent if the competition between the Rashba and
Dresselhaus terms generates small coupling constants $\mu_0$ and
$\mu_1$, regardless of the overall strength of the spin-orbit
interaction. This is the case, for example, in the vicinity of the
special symmetry points where $\mu_0$ and $\mu_1$ vanish. We
should reiterate here that the diffusion approximation may capture
spin-charge coupled dynamics only qualitatively if the spin-charge
coupling is strong. To obtain a quantitative theory in this case,
one should study  the dynamics of the matrix distribution
functions in the Boltzmann equation.

\appendix

\section{Solution of the spin diffusion equation in the limit of small
$k_-$ wave-numbers} \label{AppA}

The secular equation $\Delta(s, {\bf k})=0$ has simple analytical
solutions if one of the wave-vector components $k_-$ or $k_+$
vanishes. The asymptotic real time and real space spin dynamics is
controlled by the minima of the dispersion curves $i\omega_j({\bf
k})$ that are either at zero momentum, or at a finite value along
the $k_+$ or $k_-$ directions. To determine this asymptotic
behavior, we need the solution of the secular equation in the
vicinity of the minima. Assuming that $k_-$ is small, we obtain
for an arbitrary value of $k_+$ the following approximate
solutions of the secular equation \beqa
s_1({\bf k}) ~&=& \frac{\gamma_R\gamma_D}{2}-\xi_1(k_+)~k_-^2+{\cal O}(k_-^4), \label{appsis} \\
s_{2,3}({\bf k}) &=&-\frac12-\frac{\gamma_R\gamma_D}{4}\pm\frac12
\sqrt{\left(1-\frac{\gamma_R\gamma_D}{2}\right)^2+4k_+^2(\gamma_R+\gamma_D)^2}
-\xi_{2,3}(k_+)~k_-^2+{\cal O}(k_-^4), \nonumber \eeqa where the
coefficients of the quadratic contributions in $k_-$ are \beqa
\xi_1(k_+) ~&=&
\frac{(\gamma_R-\gamma_D)^2}{\frac{(\gamma_R+\gamma_D)^2}{\gamma_R\gamma_D}~k_+^2-
\left(1+\frac{\gamma_R\gamma_D}{2}\right)}, \label{appxis} \\
\xi_{2,3}(k_+)
&=&-\frac{\xi_1(k_+)}{2}\mp\frac{(\gamma_R-\gamma_D)^2+
\xi_1(k_+)\left(\frac12+\frac{3}{4}\gamma_R\gamma_D\right)}
{\sqrt{\left(1-\frac{\gamma_R\gamma_D}{2}\right)+4k_+^2(\gamma_R+\gamma_D)^2}}.
\nonumber \eeqa Similar expressions can be obtained for
wave-vectors with small $k_+$ and arbitrary $k_-$ components using
the correspondence rules $s_1 \leftrightarrow s_2$, $k_+
\leftrightarrow k_-$ and $\gamma_D \rightarrow -\gamma_D$. We
notice that this quadratic approximation is valid away from level
crossing points. For example, the denominator of $\xi_1$ vanishes
for a set of parameters corresponding to a degeneracy point for
the $i\omega_1$ and $i\omega_2$ modes.

The quadratic expansion around the ${\bf k}=0$ minimum of the
$i\omega_1$ mode has the form \beq i\omega_1({\bf k}) =
1-\frac{\gamma_R\gamma_D}{2} + k_+^2 + C_0^{(1)}(\gamma_R,
\gamma_D)~k_-^2+{\cal O}(k^4), \label{appw10} \eeq with \beq
C_0^{(1)}(\gamma_R, \gamma_D) =
1-\frac{(\gamma_R-\gamma_D)^2}{1+\frac{\gamma_R\gamma_D}{2}}.
\label{appC10} \eeq This zero momentum minimum exists provided
$C_0^{(1)}(\gamma_R, \gamma_D)>0$, i.e. for spin-orbit coupling
parameters corresponding to the red and blue zones in Fig.
\ref{FIGsr1}, but it does not contribute to the dynamics along the
line with equal Rashba and Dresselhaus couplings,
$\gamma_R=\gamma_D$. Similarly, we obtain for the zero momentum
minimum of the $i\omega_2$ mode that exists within the red and
yellow regions in Fig. \ref{FIGsr1} the expression \beq
i\omega_2({\bf k}) = 1+\frac{\gamma_R\gamma_D}{2} + k_-^2 +
C_0^{(2)}(\gamma_R, \gamma_D)~k_+^2+{\cal O}(k^4), \label{appw20}
\eeq with \beq C_0^{(2)}(\gamma_R, \gamma_D) =
1-\frac{(\gamma_R+\gamma_D)^2}{1-\frac{\gamma_R\gamma_D}{2}}. \eeq

For the finite momentum minimum of the $i\omega_2$ mode at $k_+^0$
that exists within the blue and green zones (see Fig.
\ref{FIGsr1}) we obtain from equations (\ref{appsis}) and
(\ref{appxis}), \beq i\omega_2({\bf k}) \approx \frac{3}{2}
+\frac{\gamma_R\gamma_D}{4}-\frac{\left(1-\frac{\gamma_R\gamma_D}{2}\right)^2}
{4(\gamma_R+\gamma_D)^2}-\frac{(\gamma_R+\gamma_D)^2}{4} +
B_{k_+^0}^{(2)}~(k_+-k_+^0)^2+ C_{k_+^0}^{(2)}~k_-^2,
\label{appw2k} \eeq where the coefficients of the quadratic terms
are \beqa B_{k_+^0}^{(2)}(\gamma_R,\gamma_D) &=&
1-\frac{\left(1-\frac{\gamma_R\gamma_D}{2}\right)^2}{(\gamma_R+\gamma_D)^4},
\label{appB2C2} \\
C_{k_+^0}^{(2)}(\gamma_R,\gamma_D) &=&
1-\frac{(\gamma_R-\gamma_D)^2}
{(\gamma_R+\gamma_D)^2}~\frac{\gamma_R^2+\gamma_D^2+\frac{5}{2}\gamma_R\gamma_D-1}
{\gamma_R^2+\gamma_D^2+\frac{1}{2}\gamma_R\gamma_D-1}. \nonumber
\eeqa Similar expressions can be obtained for the finite momentum
minimum of the $i\omega_1$ mode, corresponding to the yellow and
green regions in Fig. \ref{FIGsr1}, using the correspondence
rules: $i\omega_2\rightarrow i\omega_1$, $k_+\leftrightarrow k_-$,
and $\gamma_D\rightarrow -\gamma_D$.

\section{Explicit expressions of the diffusion matrix in the presence
of spin-charge coupling}\label{AppB}

The diffusion matrix propagator for pure Rashba or pure
Dresselhaus spin-orbit interaction can be obtained by inverting
the polarizability matrix from Eq.~(\ref{Dijm1}), after setting
$\gamma_D=0$ and $\mu_0=0$ (for the Rashba case) or $\gamma_R=0$
and $\mu_1=0$ (for Dresselhaus spin-orbit interaction). We obtain

\beq \hat{D}(\omega, {\bf k}) = \left(\begin{array}{cccc}
\frac{s}{(s-s_0)(s-s_1)} & \frac{ig\mu_l k_y}{(s-s_0)(s-s_1)} & \frac{-ig\mu_l k_x}{(s-s_0)(s-s_1)} & 0 \\
\frac{ig\mu_lk_y}{(s-s_0)(s-s_1)}&\frac{s(s^2-1)-(s-1)\gamma^2k_y^2+(s+1)g^2\mu_l^2k_x^2}{\Delta(s,{\bf
k})} &\frac{k_xk_y((s-1)\gamma^2+(s+1)g^2\mu_l^2)}{\Delta(s,{\bf
k})}&\frac{i\gamma k_x}{(s-s_2)(s-s_3)} \\
\frac{-ig\mu_lk_x}{(s-s_0)(s-s_1)}&\frac{s(s^2-1)-(s-1)\gamma^2k_y^2+(s+1)g^2\mu_l^2k_x^2}{\Delta(s,{\bf
k})}
&\frac{s(s^2-1)-(s-1)\gamma^2k_x^2+(s+1)g^2\mu_l^2k_y^2}{\Delta(s,{\bf
k})}&\frac{i\gamma k_y}{(s-s_2)(s-s_3)} \\
0&\frac{-i\gamma k_x}{(s-s_2)(s-s_3)}&\frac{-i\gamma
k_y}{(s-s_2)(s-s_3)}&\frac{s}{(s-s_2)(s-s_3)}
\end{array}
\right).                  \label{AppDij} \eeq where  $\gamma=\pm
2$ for the Rashba interaction and $\gamma\in[-2,~2]$ for the
Dresselhaus case, the spin-charge coupling parameter is given by
Eq. (\ref{mus}), and $\Delta(s, {\bf
k})=(s-s_0)(s-s_1)(s-s_2)(s-s_3)$ is the determinant of the
polarizability matrix. The solutions of the secular equation
$\Delta=0$ are $s_j({\bf k})=-i\omega_j({\bf k})+1+k^2$ with the
dispersion relations for the modes $i\omega_j$ given by
Eq.~(\ref{s14}).

For a general set of spin-orbit interaction parameters, one can
easily invert the matrix in Eq.~(\ref{calDr}) and obtain the
diffuson matrix for wave vectors oriented along the '+' and $'-'$
axes of the rotated coordinate system. Explicitly, for the $k_+$
direction we have \beq \hat{D}(\omega, k_+) =
\left(\begin{array}{cccc}
\frac{s-\frac{\gamma_R\gamma_D}{2}}{(s-s_0)(s-s_1)} & 0 & \frac{-ig(\mu_0+\mu_1) k_+}{(s-s_0)(s-s_1)} & 0 \\
0&\frac{s+1}{(s-s_2)(s-s_3)}&0&\frac{-i(\gamma_R+\gamma_D)k_+}{(s-s_2)(s-s_3)} \\
\frac{-ig(\mu_0+\mu_1)k_+}{(s-s_0)(s-s_1)}&0&\frac{s-1}{(s-s_0)(s-s_1)}&0 \\
0&\frac{i(\gamma_R+\gamma_D)k_+}{(s-s_2)(s-s_3)}&0&\frac{s+\frac{\gamma_R\gamma_D}{2}}{(s-s_2)(s-s_3)}
\end{array}
\right).                  \label{AppDwkplus} \eeq where
$s_j=-i\omega_j+1+k_+^2$ represent the four modes described by
equation (\ref{iwk14}). Notice that the charge couples only with
the $S_-$ spin component, i.e. with the in-plane component
perpendicular to the wave-vector. A similar expression can be
obtained for the $k_-$ direction using the standard correspondence
rules $'+'\leftrightarrow '-'$ and $\gamma_D\rightarrow-\gamma_D$.

\end{document}